\documentclass[english,british,aps,prx,twocolumn,floatfix,letterpaper,superscriptaddress,amssymb,eprint]{revtex4-2}
\usepackage[T1]{fontenc}
\usepackage[utf8]{inputenc}
\usepackage{babel}
\usepackage{verbatim}
\usepackage{amsmath}
\usepackage{amssymb}
\usepackage{graphicx}
\usepackage[unicode=true,pdfusetitle,
 bookmarks=true,bookmarksnumbered=false,bookmarksopen=false,
 breaklinks=false,pdfborder={0 0 1},backref=false,colorlinks=false]
 {hyperref}

\makeatletter
\usepackage{braket}
\usepackage{lmodern}
\hypersetup{colorlinks,allcolors=blue,breaklinks=true}
\allowdisplaybreaks[1]
\DeclareUnicodeCharacter{2009}{\,}

\makeatother

\begin{document}
\global\long\def\ii{\mathrm{i}}%
\global\long\def\id{\mathrm{d}}%
\foreignlanguage{english}{}
\global\long\def\k#1{\Ket{#1}}%
\foreignlanguage{english}{}
\global\long\def\b#1{\Bra{#1}}%
\foreignlanguage{english}{}
\global\long\def\bk#1{\Braket{#1}}%
\foreignlanguage{english}{}
\global\long\def\KK#1{\left\Vert #1\right\rangle }%
\foreignlanguage{english}{}
\global\long\def\BB#1{\left\langle #1\right\Vert }%

\global\long\def\GFGR{\Gamma_{\textrm{FGR}}}%
\global\long\def\tH{t_{\textrm{H}}}%
\global\long\def\Ginb{\Gamma^{\textrm{b}}}%

\title{Quantum simulation of the microscopic to macroscopic crossover using
superconducting quantum impurities}
\author{\selectlanguage{english}%
Amir Burshtein}
\email{burshtein2@mail.tau.ac.il}

\affiliation{\selectlanguage{english}%
Raymond and Beverly Sackler School of Physics and Astronomy, Tel Aviv
University, Tel Aviv 6997801, Israel}
\author{\selectlanguage{english}%
Moshe Goldstein}
\affiliation{\selectlanguage{english}%
Raymond and Beverly Sackler School of Physics and Astronomy, Tel Aviv
University, Tel Aviv 6997801, Israel}
\selectlanguage{british}%
\begin{abstract}
Despite being a pillar of quantum mechanics, little attention has
been paid to the onset of the Fermi golden rule as a discrete microscopic
bath of modes approaches the macroscopic thermodynamic limit and forms
a continuum. Motivated by recent experiments in circuit quantum electrodynamics,
we tackle this question through the lens of single-photon decay in
a finite transmission line coupled to a qubit (``quantum impurity'').
We consider a single-photon state, coupled via the nonlinear impurity
to several baths formed by multi-photon states with different number
of photons, which are inherently discrete due to the finite length of
the line. We focus on the late-time dynamics of the single-photon,
and uncover the conditions under which the photon decoherence rate
approaches the decay rate predicted by the Fermi golden rule. We show
that it is necessary to keep a small but finite escape rate (unrelated
to the impurity) for each single-photon mode to obtain a finite long-time
inelastic decay rate. We analyze the contribution of the baths formed by many-body
states with different number of photons, and illustrate how the decay
rate induced by some bath of $n$ photon states is enhanced by the
presence of other baths of $m\neq n$ photon states, highlighting
the contribution of cascade photon decay processes. Our formalism
could be used to analyze recent experiments in superconducting circuits.
\end{abstract}
\maketitle

\section{\label{sec:intro}Introduction}

Living up to the second part of its name, the Fermi golden rule (FGR)
has been an essential element in the quantum theorist's toolbox for
nearly a century \citep{dirac_quantum_1997,fermi_nuclear_1950}. As
the FGR describes the decay of a quantum state into a macroscopic
bath formed by a continuum of modes, it is natural to inquire about
what happens when the bath becomes discrete. An immediate consequence
of a discrete microscopic bath is the presence of periodic revival
processes leading to deviations from exponential decay, which have
been long reported theoretically \citep{stey_decay_1972,eberly_periodic_1980,milonni_exponential_1983,zhang_fermis_2016},
and later also experimentally, in nuclear decay \citep{kelkar_hidden_2004},
light-matter interaction \citep{rothe_violation_2006,litvinov_observation_2008},
and circuit quantum acoustodynamics \citep{andersson_non-exponential_2019}.
However, a closely-related question has been addressed directly only
recently \citep{micklitz_emergence_2022} --- how does the FGR emerge
as the bath approaches the thermodynamic limit? Important motivations
to this question have been studies of Fock-space localization \citep{altshuler_quasiparticle_1997}
and many-body localization \citep{basko_metalinsulator_2006,dalessio_quantum_2016,abanin_colloquium_2019},
which monitor the dynamics of some initial quantum state in a large
interacting system, as well as works on thermalization in Floquet
systems \citep{seetharam_absence_2018,morningstar_universality_2023},
weak breaking of integrability and the onset of chaos \citep{znidaric_weak_2020,brenes_eigenstate_2020,bulchandani_onset_2022},
and quantum scars \citep{bernien_probing_2017}.

Experimentally, the process of approaching the thermodynamic limit
could be directly realized in quantum simulators. A particularly convenient
arena is provided by superconducting circuits, which allow for natural
implementation of quantum impurity models, comprised of a Josephson
array coupled to some nonlinearity imposed by a qubit. It was demonstrated
that a single-photon propagating in the Josephson array could decay
into multi-photon states with a high probability \citep{kuzmin_inelastic_2021},
due to superstrong coupling between the Josephson array and the qubit
\citep{kuzmin_superstrong_2019,puertas_martinez_tunable_2019}, and
that the decay rates of those single-photons could be used as experimental
and theoretical probes that elucidate the dynamics of the impurity
\citep{le_hur_kondo_2012,goldstein_inelastic_2013,gheeraert_particle_2018,houzet_critical_2020,burshtein_photon-instanton_2021,leger_revealing_2023,kuzmin_observation_2023,burshtein_inelastic_2024,houzet_microwave_2024,remez_bloch_2024, burshtein_numerical_2024}.
While the Josephson arrays are very long and support a large number
($N\sim10^{4}$) of single-photon states, they are inevitably finite,
such that the photon spectrum is discrete rather than continuous.
Indeed, Ref. \citep{mehta_down-conversion_2023} demonstrated that
shortening the transmission line gives rise to avoided-crossings in
the single-photon spectroscopy picture, due to level repulsions between
the single-photon state and the discrete set of many-photon states.
The mode spacing of $n$-photon states scales as $\Delta_{n}\sim\ell^{-n}$,
where $\ell$ is the length of the Josephson array, and could be tuned
to investigate the finite-size effects; relatedly, Ref. \citep{fraudet_direct_2024}
demonstrated how the finite length of the array could be used to tailor
the decay of a single-photon into a specific outgoing multi-photon
state.

With experiments already exploring the effects of the finite length
of the array, a theoretical analysis is in demand. Some work has already
been performed; finite-size effects in circuit quantum electrodynamics
were studied \citep{giacomelli_emergent_2024} in the context of the
Schmid-Bulgadaev quantum phase transition \citep{schmid_diffusion_1983,bulgadaev_phase_1984}
and dual Shapiro steps \citep{borletto_circuit_2024}. More specifically,
in the context of single-photon scattering, a phenomenological theory
was proposed in Ref. \citep{mehta_theory_2022} to reproduce the spectroscopy
probed in Ref. \citep{mehta_down-conversion_2023}. Yet, the decay
rates of the single-photons were not addressed, nor was the thermodynamic
limit of a semi-infinite Josephson array.

\begin{figure*}
\begin{centering}
\includegraphics[width=0.99\textwidth]{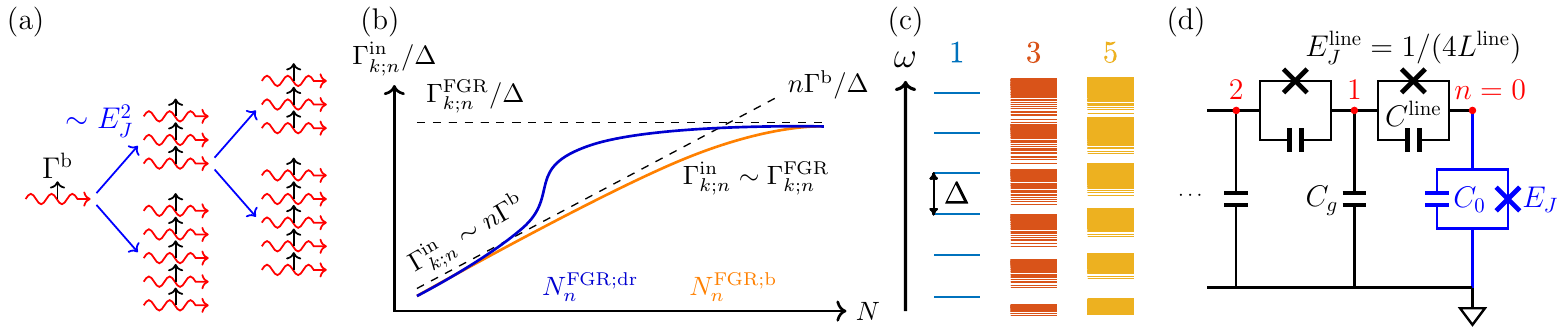}
\par\end{centering}
\caption{\label{fig:intro_fig}(a) Each photon is coupled to multiple many-body
states via the impurity, and may also decay into the environment with
a rate $\protect\Ginb$. (b) Sketch of the main result, displaying
the decay rate $\Gamma_{k;n}^{\textrm{in}}$ of some mode $k$ induced
by a bath of $n$-photon states with (blue) and without (orange) the
dressing due to cascade decay processes into lower-frequency modes,
as a function of the system size $N$. The rates are scaled by the
single-photon mode spacing, $\Delta\sim1/N$. At small $N$, $\Gamma_{k;n}^{\textrm{in}}$
is bounded by the bare broadening of the bath modes, $n\protect\Ginb$.
The decay rate approaches the rate predicted by the FGR, $\Gamma_{k;n}^{\textrm{FGR}}$,
at large enough $N$, where $\Gamma_{k;n}^{\textrm{FGR}}<n\protect\Ginb$.
The dressing induced by the baths of $m\protect\neq n$-photon states
reduces the crossover scale for the array size, $N_{n}^{\textrm{FGR;dr}}$
as compared to the scale $N_{n}^{\textrm{FGR;b}}$ in their absence,
and allows for $\Gamma_{k;n}^{\textrm{in}}>n\protect\Ginb$ in the
crossover region. (c) Energies of 1-, 3-, and 5-photon states. The
multi-photon levels are spread out by disorder and the curvature of
the dispersion relation. (d)~Quantum impurity model implemented by
a Josephson array terminated with a Cooper-pair box.}
\end{figure*}

In this work, we show how the FGR emerges in the inelastic scattering
of single-photons off an impurity in a finite Josephson array. We
consider an interacting model of a (nearly) quadratic bosonic transmission
line, supporting photon eigenstates, terminated by a nonlinear impurity,
depicted in Fig.~\ref{fig:intro_fig}(d). The photons are coupled to
one another via wave-mixing terms provided by the impurity, such that
each single-photon state is coupled to many multi-photon states; states
with different photon numbers thus behave as different baths. We focus
on the late-time behavior of the survival probability of the initial
state, and show that, if the array size is finite, one must keep a
finite escape rate (unrelated to the impurity) of the single-photon states, unrelated to
the nonlinearities in the impurity, to obtain a finite long-time inelastic
rate. We uncover the conditions under which the inelastic decay rate induced
by an $n$-photon bath approaches the rate predicted by the FGR for
the corresponding infinite system, at large enough mode frequencies
and array sizes. Applying a self-consistent approach, we show that
the crossover of the $n$-photon bath into the thermodynamic regime,
where its induced decay rate is given by the FGR, is accelerated by
the presence of baths of $m\neq n$-photon states, thanks to cascade
processes depicted in Fig.~\ref{fig:intro_fig}(a). We further show that,
in the crossover region of the $n$-photon bath, the $m\neq n$-photon
baths [see Fig.~\ref{fig:intro_fig}(c)] enhance the decay rate into the $n$-photon bath, allowing for
values which would not have been possible in their absence. Our main
results are sketched in Fig.~\ref{fig:intro_fig}(b), demonstrating the
crossover of the $n$-photon bath rate from the microscopic regime,
where it is dominated by the external decoherence rate of the single-photon
states, into the macroscopic regime, where it approaches the FGR rate.
Fig.~\ref{fig:intro_fig}(b) further illustrates the importance of the
cascade decay processes, which enhance the $n$-photon bath rate in
the crossover region and accelerate the convergence to the thermodynamic
limit.

In this work we consider a Cooper-pair box impurity, where many multi-photon
baths provide significant contributions to the total decay rate. Yet,
our formalism could be easily adapted to explicitly address the finite-size
effects on the decay rate of single-photons scattering off any type
of impurity, and in particular could be used to analyze the experiment
in Ref. \citep{mehta_down-conversion_2023}, which considered a fluxonium
impurity \citep{manucharyan_fluxonium_2009}, and also applies to
nonlinearities induced by weak bulk perturbations \citep{kuzmin_quantum_2019,bard_decay_2018,wu_theory_2019,houzet_microwave_2019}.
Similarly, we could also use our results to evaluate decoherence times
of qubits coupled to a finite multimode resonator, which could be
useful for quantum information processing applications.

The rest of the paper is organized as follows. In Section \ref{sec:toy-model},
we discuss an exactly-solvable quadratic toy model of a resonant level
coupled to an equidistant discrete bath. Though cascade decay processes
are absent in this case, it nevertheless can teach us about some of
the relevant effects. In particular, we show that an external broadening
of the bath modes must be introduced to get a finite decay rate of
the resonant level in the long-time limit, and find the conditions
under which the decay rate approaches the FGR value. Then, in Section
\ref{sec:CPB}, we consider the interacting model of a Josephson array
coupled to a Cooper-pair box. We find the propagators of the single-photon
modes by applying a self-consistent approach, which accounts for cascade
processes into low-frequency modes. We calculate the FGR rates induced
by each bath of $2n+1$-photon states, and evaluate numerically the
decay rates at finite array size, demonstrating the conditions for
the microscopic to macroscopic crossover. We conclude in Section \ref{sec:conclusions}.
The technical details of the evaluation of the FGR decay rates of
the baths are relegated to Appendix \ref{app:FGR_2n+1_calc}, and
details regarding the numerical evaluation of the decay rates of the
interacting system are discussed in Appendix \ref{app:numerical_details}.

\section{\label{sec:toy-model}Toy model --- resonant level coupled to an
equispaced bath}

In this Section we consider a quadratic model, which cannot have cascade
decay processes, yet can inform us some useful lessons. We study the
Hamiltonian of the resonant level model,
\begin{equation}
\mathcal{H}=\frac{\epsilon_{d}}{2}d^{\dagger}d+\sum_{n}\epsilon_{n}a_{n}^{\dagger}a_{n}+\sum_{n}\left(g_{n}da_{n}^{\dagger}+g_{n}^{*}d^{\dagger}a_{n}\right),\label{eq:H_toy}
\end{equation}
where $d,a_{n}$ are annihilation operators (either fermionic or bosonic)
of the resonant level at $\epsilon_{d}$ and the bath modes at $\epsilon_{n}$,
respectively, and $g_{n}$ are the coupling coefficients. Our goal
is to obtain the decay rate $\Gamma$ of the resonant level, determined
from the late-time behavior of the survival probability, $\mathcal{P}_{d}\left(t\right)=\left|\bk{d\left(t\right)d^{\dagger}\left(0\right)}\right|^{2}\sim e^{-\Gamma t}$
as $t\rightarrow\infty$. The survival probability can be extracted
from the time-ordered propagator of the resonant mode, $\mathcal{P}_{d}\left(t\right)=\left|G_{d}\left(t\right)\right|^{2}$,
where $G_{d}\left(t\right)=-\ii\left\langle \mathcal{T}d\left(t\right)d^{\dagger}\left(0\right)\right\rangle $,
and $\mathcal{T}$ stands for time-ordering. We calculate expectation
values with respect to the vacuum state $\k 0$ which is annihilated by $d,a_{n}$
(that is, $d\k 0=a_{n}\k 0=0$), though generalization to expectation
values in general excited states is straightforward. Since the toy
model in Eq.~(\ref{eq:H_toy}) is quadratic, an exact solution for
the time-ordered propagator is readily available in the frequency
domain,
\begin{equation}
G_{d}\left(\omega\right)=-\ii\int_{-\infty}^{\infty}\id te^{\ii\omega t}\bk{\mathcal{T}d\left(t\right)d^{\dagger}\left(0\right)}=\frac{1}{\omega-\epsilon_{d}-\Sigma_{d}\left(\omega\right)},
\end{equation}
where $\Sigma_{d}\left(\omega\right)=\sum_{n}\left|g_{n}\right|^{2}/\left(\omega-\epsilon_{n}\right)$
is the spectral function of the bath. From the inverse Fourier transform
of $G_{d}\left(\omega\right)$, we find
\begin{equation}
\mathcal{P}_{d}\left(t\right)=\left|\int_{-\infty}^{\infty}\frac{\id\omega}{2\pi}\frac{e^{-\ii\omega t}}{\omega-\epsilon_{d}-\Sigma_{d}\left(\omega\right)}\right|^{2}=\left|\sum_{z_{p}}r_{p}e^{-\ii z_{p}t}\right|^{2},
\end{equation}
where $z_{p}$ are solutions to $z-\epsilon_{d}-\Sigma_{d}\left(z\right)=0$
with $\mathrm{Im}z_{p}<0$, and $r_{p}=\mathrm{Res}_{z=z_{p}}1/\left(z-\epsilon_{d}-\Sigma_{d}\left(z\right)\right)$.
At $t\rightarrow\infty$, the survival probability is dominated by
the pole with the smallest negative imaginary part (in absolute value),
such that $\Gamma=-2\min_{z_{p}}\left|\mathrm{Im}z_{p}\right|$.

First, consider the textbook example of a bath composed of a continuum
of modes with a constant density of states $\nu$, spread across a
bandwidth $2D$, and with a uniform coupling $g$. The Hamiltonian reads
\begin{align}
\mathcal{H}= & \frac{\epsilon_{d}}{2}d^{\dagger}d+\nu\int_{-D}^{D}\mathrm{d\varepsilon}a^{\dagger}\left(\varepsilon\right)a\left(\varepsilon\right)\nonumber \\
 & +\nu g\int_{-D}^{D}\mathrm{d}\varepsilon\left(da^{\dagger}\left(\varepsilon\right)+d^{\dagger}a\left(\varepsilon\right)\right),
\end{align}
and the spectral function of the bath in the complex plane is given
by

\begin{align}
\Sigma_{d}\left(z=\omega+\ii\chi\right) & =\nu\left|g\right|^{2}\lim_{\eta\rightarrow0^{+}}\log\frac{z+D+\ii\eta}{z-D+\ii\eta}\nonumber \\
 & \hspace{-2cm}=\nu\left|g\right|^{2}\log\left(-\frac{D^{2}-\omega^{2}-\chi^{2}+2\ii\chi D}{\left(\omega-D\right)^{2}+\chi^{2}}\right).\label{eq:toy_model_sigma_continuum}
\end{align}
Taking the limit of a large bandwidth, $D\gg\nu\left|g\right|^{2}$
and $D\gg\epsilon_{d}$, we find that $z=\epsilon_{d}-\ii\pi\nu\left|g\right|^{2}$
is a solution to $z-\epsilon_{d}-\Sigma_{d}\left(z\right)=0$, up
to corrections of order $\nu\left|g\right|^{2}/D$ and $\epsilon_{d}/D$,
and we recover the textbook result, $\Gamma=\GFGR=-2\mathrm{Im}\Sigma_{d}\left(\omega=\epsilon_{d}\right)=2\pi\nu\left|g\right|^{2}$.

Let us now we consider a discrete bath of uniformly-coupled ($g_{n}=g$),
evenly-spaced modes, $\epsilon_{n}=\epsilon_{d}+\delta_{\min}+n\Delta$,
where $n\in\mathbb{Z}$, $\Delta$ is the mode spacing, and $\delta_{\min}\in\left[-\Delta/2,\Delta/2\right]$
is the distance from the resonant level at $\epsilon_{d}$ to the
closest bath mode. We seek the conditions under which the decay rate
approaches the FGR rate predicted from the corresponding continuum
model, $\GFGR=2\pi\left|g\right|^{2}/\Delta$.

Again, we are looking for solutions to the equation $z-\epsilon_{d}-\Sigma_{d}\left(z\right)=0$,
with $\Sigma_{d}\left(z\right)=\sum_{n}\left|g\right|^{2}/\left(z-\epsilon_{n}\right)$.
As in Eq.~(\ref{eq:toy_model_sigma_continuum}), we introduce an infinitesimal
imaginary part to the bath energies, $\epsilon_{n}\rightarrow\epsilon_{n}-\ii\eta/2$,
and in the end take the limit $\eta\rightarrow0^{+}$. From the imaginary
part of the equation for the poles, we find
\begin{equation}
\chi=-\sum_{n}\frac{\left|g\right|^{2}\left(\eta/2+\chi\right)}{\left(\omega-\epsilon_{n}\right)^{2}+\left(\eta/2+\chi\right)^{2}},\label{eq:chi_imaginary_discrete_uniform}
\end{equation}
where, as before, $z=\omega+\ii\chi$. Recall that we are looking
for poles with a negative imaginary part, $\chi<0$. Therefore, it
must hold that $\eta/2+\chi>0$, such that $\Gamma=-2\chi<\eta$.
We then find that $\Gamma$ vanishes upon taking the limit $\eta\rightarrow0^{+}$,
regardless of the coupling strength $g$ or the mode spacing $\Delta$.
The decay rate vanishes due to the manifestly discrete nature of the
bath. Taking the limit $\Delta\rightarrow0$ prior to $\eta\rightarrow0^{+}$
would lead to $\Sigma_{d}\left(z\right)=\ii\pi\left|g\right|^{2}/\Delta$
at any $\mathrm{Im}z<0$, and the FGR rate would be recovered. On
the contrary, taking the limit $\eta\rightarrow0^{+}$ while keeping
a small but finite mode spacing $\Delta$ retains the discrete structure,
such that revival processes occur at times $t=m\tH$, where $m=1,2,\ldots$,
and $\tH=2\pi/\Delta$ is the Heisenberg time of the bath. These periodic
revivals are explicitly present in the following analytic expression
for the propagator of the resonant mode in the time domain \citep{stey_decay_1972,lefebvre_memory_1974},
\begin{align}
G_{d}\left(t\right) & =e^{-\GFGR t/2}+\sum_{m=1}^{\infty}e^{-\GFGR\left(t-m\tH\right)/2}e^{\ii\delta_{\min}m\tH}\nonumber \\
 & \hspace{-1.2cm}\times\Theta\left(t-m\tH\right)\sum_{l=0}^{m-1}{m-1 \choose l}\frac{\left[-\left(\left(t-m\tH\right)\GFGR\right)\right]^{m-l}}{\left(m-l\right)!},\label{eq:toy_model_anal_eta=00003D0}
\end{align}
where $\Theta\left(x\right)$ is the Heaviside function. This implies
that the limit $\lim_{t\rightarrow\infty}\mathcal{P}_{d}\left(t\right)$
does not exist.

In order to obtain a finite decay rate, it is necessary to keep finite
external broadenings for the bath modes, $\epsilon_{n}\rightarrow\epsilon_{n}-\ii\eta/2$.
Formally, $\eta$ could be obtained by coupling the bath modes $a_{n}$
to an additional continuous bath, and tracing over the degrees of
freedom of the continuous bath. The self-energy becomes
\begin{equation}
\Sigma_{d}\left(z\right)=\frac{\GFGR}{2}\cot\left(\frac{\pi}{\Delta}\left(z-\delta_{\min}+\ii\eta/2\right)\right),
\end{equation}
and solving for the equation $z-\epsilon_{d}-\Sigma_{d}\left(z\right)=0$
numerically yields the decay rate $\Gamma$. The propagator in the
time domain becomes
\begin{align}
G_{d}\left(t\right) & =e^{-\GFGR t/2}\nonumber \\
 & \hspace{-1cm}+\sum_{m=1}^{\infty}e^{-\eta m\tH/2}e^{-\GFGR\left(t-m\tH\right)/2}e^{\ii\delta_{\min}m\tH}\nonumber \\
 & \hspace{-1cm}\times\Theta\left(t-m\tH\right)\sum_{l=0}^{m-1}{m-1 \choose l}\frac{\left[-\left(\left(t-m\tH\right)\GFGR\right)\right]^{m-l}}{\left(m-l\right)!},\label{eq:toy_model_anal}
\end{align}
such that the $m$th revival process is attenuated by a factor
$e^{-\eta m\tH/2}$. From Eq.~(\ref{eq:toy_model_anal}), it is easy
to extract the late-time decay rate in two limiting cases:
\begin{equation}
\Gamma\sim\begin{cases}
\eta, & \Gamma_{\textrm{FGR}}\gg\eta,\Delta,\\
\Gamma_{\textrm{FGR}}, & \eta\gg\Gamma_{\textrm{FGR}},\Delta.
\end{cases}\label{eq:Pt_clean_summary}
\end{equation}
Note that the two limits hold for any $\eta/\Delta$ in the first
case and any $\GFGR/\Delta$ in the second case. The intuition behind
this result is clear --- approaching the thermodynamic limit, $\Delta\rightarrow0$,
the survival probability at large times is dominated by the slower
among the decay rate into the bath, $\GFGR$, and the escape rate from
the bath into the open environment, $\eta$. When the mode spacing
$\Delta$ exceeds both $\eta$ and $\GFGR$, there is no simple asymptotic
expression for $\Gamma$; revival processes must be summed coherently,
leading to a large time behavior which is sensitive to the position
of the resonant level with respect to the equidistant bath modes.

\begin{figure}
\begin{centering}
\includegraphics[width=0.99\columnwidth]{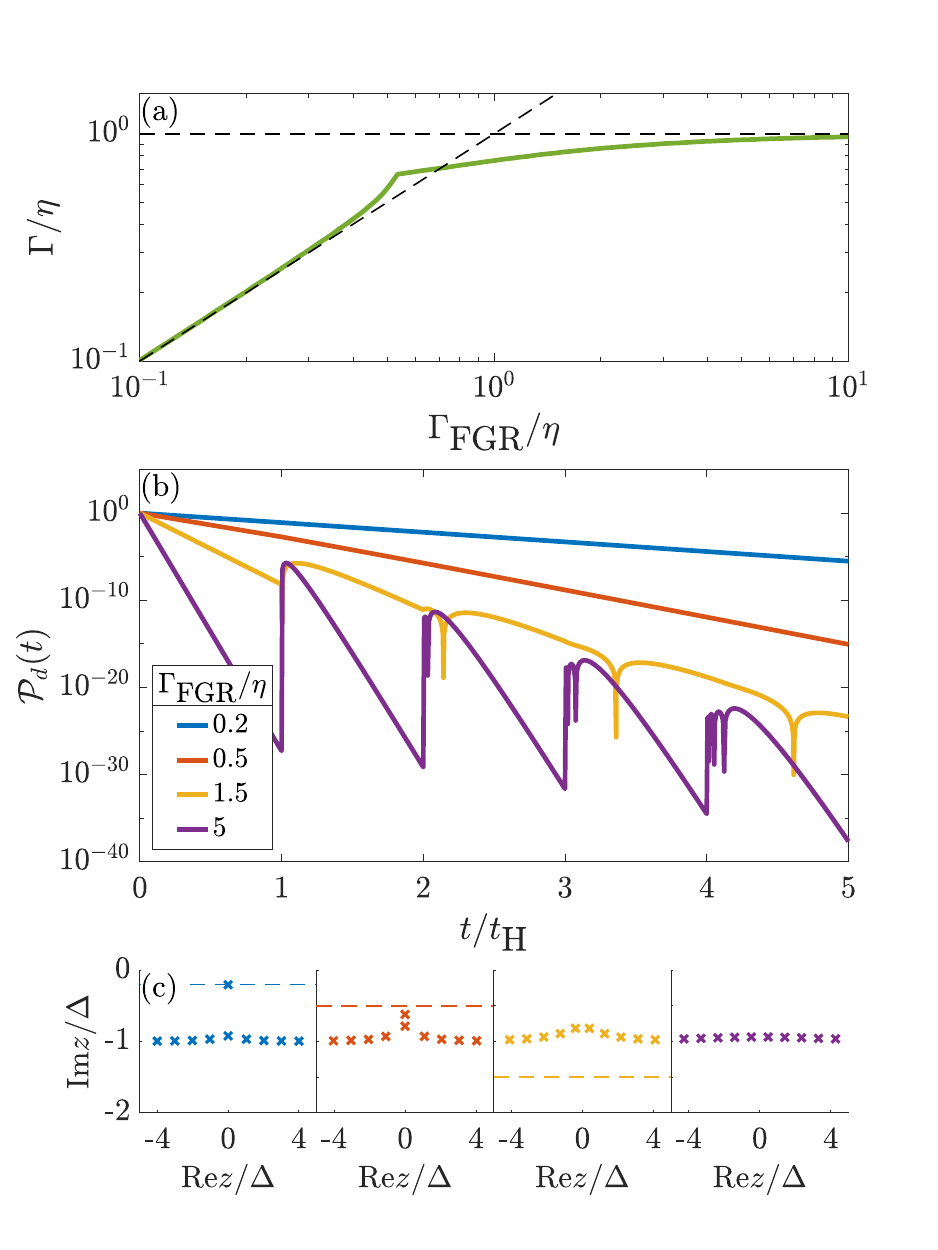}
\par\end{centering}
\caption{\label{fig:poles}Decay rate for the toy model in Eq.~(\ref{eq:H_toy})
for an equidistant bath, $\epsilon_{n}=n\Delta$ and $N\rightarrow\infty$,
with uniform coupling, $g_{n}=g$, external broadening $\eta=2\Delta$,
and $\epsilon_{d}=0$. (a) Late-time decay rate $\Gamma$ ($\mathcal{P}_{d}\left(t\rightarrow\infty\right)\sim e^{-\Gamma t}$)
as a function of $\protect\GFGR/\eta$. The dashed lines show the
asymptotes provided by Eq.~(\ref{eq:Pt_clean_summary}). (b) Survival
probability, $\mathcal{P}_{d}\left(t\right)=\left|G_{d}\left(t\right)\right|^{2}$,
given by Eq.~(\ref{eq:toy_model_anal}). (c) The poles of the propagator
$G_{d}\left(z\right)$ in the complex plane. The dashed lines are
at $\mathrm{Im}z=-\protect\GFGR/2$.}
\end{figure}

The survival probability, decay rate, and poles of $G_{d}\left(z\right)=1/\left(z-\epsilon_{d}-\Sigma_{d}\left(z\right)\right)$
for the uniform, equidistant bath with $\epsilon_{d}=\delta_{\min}=0$
and $\eta/\Delta=2$ are displayed in Fig.~\ref{fig:poles}. The survival
probability clearly illustrates the revival processes occuring
at Heisenberg times $t=m\tH$ when the escape rate into the bath,
$\GFGR$, exceeds the external bath broadening $\eta$. In the complex
plane, $G_{d}\left(z\right)$ always displays a set of poles at $\mathrm{Im}z=-\eta/2$, separated apart by $\Delta$ along the real axis, and there is a pole
at $\mathrm{Im}z=-\GFGR/2$ when $\GFGR\ll\eta$.

Finally, let us note that for a bath with general coupling coefficients
$g_{n}$, mode energies $\epsilon_{n}$, and external broadenings
$\eta_{n}$, the decay rate is always bounded by the maximal external
broadening of the bath modes. Eq.~(\ref{eq:chi_imaginary_discrete_uniform})
generalizes to
\begin{equation}
\chi=-\sum_{n}\frac{g_{n}^{2}\left(\eta_{n}/2+\chi\right)}{\left(\omega-\epsilon_{n}\right)^{2}+\left(\eta_{n}/2+\chi\right)^{2}}.\label{eq:chi_imaginary_discrete}
\end{equation}
Again, since $\chi$ must be negative, it must hold that $\eta_{n}/2+\chi>0$
at least for one of the bath modes. In other words, the late-time
decay rate is bounded by the maximal bath broadening, $\Gamma=-2\chi<\max_{n}\left\{ \eta_{n}\right\} $.

\section{\label{sec:CPB}Experimental interacting system --- Josephson array
coupled to a Cooper-pair box}

We now depart from the quadratic toy model, and turn to analyze an
interacting model of a Josephson array, implementing a nearly quadratic
transmission line, terminated by a nonlinear impurity. The Josephson
array is diagonalized by photonic modes, which are coupled to one
another via wave-mixing terms provided by the impurity, such that
each single-photon state is coupled to many multi-photon states. For
any given $n$, we refer to the set of $n$-photon states as a bath.

As before, in order to discuss the long-time decay rate, we must introduce
some broadening of the modes by an external environment. This broadening
is inherently present in any experimental system, due to loss mechanism
in the array that are unrelated to the wave-mixing terms provided
by the impurity. The origin of these mechanisms could be both internal,
such as quasiparticle poisoning \citep{martinis_energy_2009} or two-level-system
dielectric loss \citep{martinis_decoherence_2005}, as well as external,
due to the capacitive coupling of the array
to the antenna used for spectroscopy. We incorporate these loss mechanisms
into the calculation by assigning each mode some initial broadening;
note that this broadening may be directly probed in an experiment
by disconnecting the impurity and measuring the internal and external
quality factors of the system eigenmodes \citep{kuzmin_quantum_2019}.

In this Section, we derive an expression for the propagator of the
single-photon modes. We identify the contributions of the different
baths, and, following our discussion in Section \ref{sec:toy-model},
analyze the decay rate induced by each bath. Applying a self-consistent
approach, we show how the convergence of the decay rate induced by
some bath to its FGR rate is accelerated by the presence of other
baths, due to cascade decay processes into lower-frequency modes,
depicted in Fig.~\ref{fig:intro_fig}.

For concreteness, in the following, we consider a Cooper-pair box
impurity, implementing a discrete analog of the boundary sine-Gordon
model, although our formalism applies to any type of impurity, and
also to wave-mixing terms originating from weak bulk nonlinearities.

\subsection{Hamiltonian and eigenmodes}

Consider the Hamiltonian of a Josephson array coupled to a Cooper-pair
box (setting $\hbar=e=1$ henceforth), sketched in Fig.~\ref{fig:intro_fig},
\begin{align}
\mathcal{H}= & \sum_{i,j=0}^{N}2\left[\mathrm{C}^{-1}\right]_{i,j}Q_{i}Q_{j}\nonumber \\
 & -\sum_{i=1}^{N}E_{J}^{\textrm{line}}\cos\left(\phi_{i}-\phi_{i-1}\right)-E_{J}\cos\left(\phi_{0}\right),\label{eq:H_CPB}
\end{align}
where $\phi_{i}$ and $Q_{i}$ are the superconducting phase and charge
of the $i$th superconducting grain, respectively. The matrix $\mathrm{C}^{-1}$
is the inverse of the capacitance matrix $\mathrm{C}$, which is a
tridiagonal matrix with $\mathrm{C}_{0,0}=C_{0}+C^{\textrm{line}}$,
$\mathrm{C}_{i,i-1}=\mathrm{C}_{i-1,i}=-C^{\textrm{line}}$, $\mathrm{C}_{i,i}=C_{g}+2C^{\textrm{line}}$
for $1\le i\le N-1$, and $\mathrm{C}_{N,N}=C_{g}+C^{\textrm{line}}$.
The Josephson energies $E_{J}^{\textrm{line}}$ and intergrain capacitances
$C^{\textrm{line}}$ are chosen such that $E_{J}^{\textrm{line}}\gg E_{C}^{\textrm{line}}$,
where $E_{C}^{\textrm{line}}=1/\left(2C^{\textrm{line}}\right)$,
so that anharmonic effects and phase slips are suppressed in the line,
and the Josephson terms of the array may be replaced by $\left(\phi_{i}-\phi_{i-1}\right)^{2}/\left(8L^{\textrm{line}}\right)$,
with $L^{\textrm{line}}=1/\left(4E_{J}^{\textrm{line}}\right)$. Treating
the boundary cosine term as a perturbation, the quadratic part of
the Hamiltonian is diagonalized by plane waves, $\phi_{i}=\sum_{k}\phi_{k}\sin\left(ki+\delta_{k}\right)$
(where we set the grain spacing to unity), with the oscillatory behavior
$\ddot{\phi}_{k}=-\omega_{k}^{2}\phi_{k}$. The dispersion relation
reads $\omega_{k}\approx vk/\sqrt{1+\left(vk/\omega_{p}\right)^{2}}$,
with $v=1/\sqrt{L^{\textrm{line}}C_{g}}$ and $\omega_{p}=1/\sqrt{L^{\textrm{line}}C^{\textrm{line}}}$,
and, assuming $v\gg\omega_{p}$, the phase shift is given by $\tan\delta_{k}\approx-\Gamma_{0}\sqrt{1-\left(\omega_{k}/\omega_{p}\right)^{2}}/\omega_{k}$.
Here $\Gamma_{0}=1/\left(ZC_{0}\right)$ is the inverse $RC$ time
of the array and the Cooper-pair box, and $Z=\sqrt{L^{\textrm{line}}/C_{g}}$
is the impedance of the line. In the following, we denote the normalized
impedance by $z=Z/R_{Q}$, where $R_{Q}=h/\left(4e^{2}\right)=\pi/2$
is the resistance quantum. Open boundary conditions at $i=N$ lead
to the quantization condition $kN+\delta_{k}=\pi l$, where $l=0,1,\ldots,N-1$,
giving rise to the $k$-dependent mode spacing, $\Delta_{k}=\Delta\left(1-\left(\omega_{k}/\omega_{p}\right)^{2}\right)^{3/2}$,
with $\Delta=\pi v/N$.

\subsection{Self-energy and multi-photon baths}

Introducing creation and annihilation operators, $\phi_{k}=\sqrt{2z\Delta\left(1-\left(\omega_{k}/\omega_{p}\right)^{2}\right)/\omega_{k}}\left(a_{k}+a_{k}^{\dagger}\right)$,
leads to $\mathcal{H}=\mathcal{H}_{0}+\mathcal{H}_{I}$, where the
free Hamiltonian is $\mathcal{H}_{0}=\sum_{k}\omega_{k}a_{k}^{\dagger}a_{k}$,
and the interaction term reads $\mathcal{H}_{I}=-E_{J}\cos\left(\sum_{k}f_{k}\left(a_{k}+a_{k}^{\dagger}\right)\right)$,
with
\begin{equation}
f_{k}^{2}=\frac{2z\Delta}{\omega_{k}}\frac{\left(1-\left(\omega_{k}/\omega_{p}\right)^{2}\right)^{2}}{1+\left(\left(\omega_{p}/\Gamma_{0}\right)^{2}-1\right)\left(\omega_{k}/\omega_{p}\right)^{2}}.
\end{equation}
Throughout this work, we assume $E_{J}\ll E_{C}$ with $E_{C}=1/\left(2C_{0}\right)$,
such that $\mathcal{H}_{I}$ can be treated as a perturbation. Note
that $\mathcal{H}_{I}$ is relevant in an RG sense for $z<1$ \citep{gogolin_bosonization_2004},
and there is an emergent RG scale, $E_{J}^{\star}\sim\left(E_{J}/\omega_{c}^{z}\right)^{1/\left(1-z\right)}$,
where $\omega_{c}\sim\min\left\{ \omega_{p},\Gamma_{0}\right\} $
is a UV cutoff, such that perturbation theory applies only at $\omega_{k}\gg E_{J}^{\star}$.
$\mathcal{H}_{I}$ is irrelevant for $z>1$, such that $E_{J}^{\star}=0$,
and perturbation theory applies at all frequencies. In the following,
we address the perturbative regime. To second order in $E_{J}$, the
time-ordered propagator of the $k$th mode, $G_{k}\left(t\right)=-\ii\left\langle \mathcal{T}\phi_{k}\left(t\right)\phi_{k}\left(0\right)\right\rangle $,
is given in the frequency domain by
\begin{equation}
G_{k}\left(\omega\right)=\frac{4z\Delta\left(1-\left(\omega_{k}/\omega_{p}\right)^{2}\right)}{\omega^{2}-\left[\omega_{k}-\left(\Sigma_{k}\left(\omega\right)-\mathrm{Re}\Sigma_{k}\left(\omega=0\right)\right)\right]^{2}},
\end{equation}
 with the self-energy (at $T=0$) \citep{kuzmin_observation_2023}
\begin{equation}
\Sigma_{k}\left(\omega\right)=2\ii E_{J}^{2}f_{k}^{2}\int_{-\infty}^{\infty}\id te^{\ii\omega t}\left\langle \mathcal{T}\sin\left(\phi_{0}\left(t\right)\right)\sin\left(\phi_{0}\left(0\right)\right)\right\rangle _{\mathcal{H}_{0}},\label{eq:Sigma_k}
\end{equation}
where the expectation value is calculated with respect to $\mathcal{H}_{0}$.
Using the eigenmode expansion of $\phi_{0}$ and the Campbell-Baker-Hausdorff
formula, we find:
\begin{align}
\Sigma_{k}\left(\omega\right)= & 2\ii E_{J}^{2}f_{k}^{2}e^{-\sum_{k^{\prime}}f_{k^{\prime}}^{2}}\nonumber \\
 & \times\int_{-\infty}^{\infty}\id te^{\ii\omega t}\sinh\left\{ \sum_{k^{\prime}}\ii f_{k^{\prime}}^{2}\tilde{G}_{k^{\prime}}^{\left(0\right)}\left(t\right)\right\} ,\label{eq:Sigma_sinh}
\end{align}
where $\tilde{G}_{k}^{\left(0\right)}=-\ii\left\langle \mathcal{T}\left(a_{k}\left(t\right)+a_{k}^{\dagger}\left(t\right)\right)\left(a_{k}\left(0\right)+a_{k}^{\dagger}\left(0\right)\right)\right\rangle _{\mathcal{H}_{0}}$
is the free propagator of the $k$th mode (without the prefactor appearing
in $\phi_{k}$), given by
\begin{equation}
\tilde{G}_{k}^{\left(0\right)}\left(\omega\right)=\frac{2\omega_{k}}{\omega^{2}-\left(\omega_{k}-\ii\Ginb/2\right)^{2}}.
\end{equation}
Note that we introduce some finite broadening, $\Ginb$, to each of
the free propagators. Similarly to Section \ref{sec:toy-model}, $\Ginb$
could be formally obtained by coupling the free modes to a larger
environment and tracing over the degrees of freedom of the environment.
This broadening is due internal and external dissipation processes
in the array which are unrelated to the impurity at $i=0$, is always
present in an experimental environment, and can be assumed to be the
smallest energy scale in the system, several orders of magnitude below
the mode spacing $\Delta$ \citep{kuzmin_quantum_2019}.

Expanding the hyperbolic sine in Eq.~(\ref{eq:Sigma_sinh}), we find
$\Sigma_{k}\left(\omega\right)=\sum_{n=0}^{\infty}\Sigma_{k;2n+1}\left(\omega\right)$,
with
\begin{align}
\Sigma_{k;2n+1}^{\textrm{b}}\left(\omega\right)= & \frac{2\ii E_{J}^{2}f_{k}^{2}e^{-\sum_{k^{\prime}}f_{k^{\prime}}^{2}}}{\left(2n+1\right)!}\nonumber \\
 & \hspace{-1cm}\times\int_{-\infty}^{\infty}\id te^{\ii\omega t}\left[\sum_{k^{\prime}}\ii f_{k^{\prime}}^{2}\tilde{G}_{k^{\prime}}^{\left(0\right)}\left(t\right)\right]^{2n+1}.\label{eq:Sigma_2n+1_bare}
\end{align}
$\Sigma_{k;2n+1}^{\textrm{b}}\left(\omega\right)$ is the contribution
of the bath of $2n+1$ photon modes to the self-energy $\Sigma_{k}^{\textrm{b}}\left(\omega\right)$.
The superscript $\textrm{b}$ indicates that the self-energy is evaluated
using the bare propagators, $\tilde{G}_{k^{\prime}}^{\left(0\right)}\left(t\right)$,
and stands to distinguish it from the self-consistent self-energy,
to be discussed below. To understand the structure of $\Sigma_{k;2n+1}^{\textrm{b}}\left(\omega\right)$,
it is useful to expand the product and express the result in terms
of the frequency-domain bare propagators:
\begin{align}
\Sigma_{k;2n+1}^{\textrm{b}}\left(\omega\right)= & \frac{2\ii E_{J}^{2}f_{k}^{2}e^{-\sum_{k^{\prime}}f_{k^{\prime}}^{2}}}{\left(2n+1\right)!}\sum_{k_{1},\ldots,k_{2n+1}}\prod_{i=1}^{2n+1}f_{k_{i}}^{2}\nonumber \\
 & \hspace{-1cm}\times\frac{1}{\left(2\pi\right)^{2n}}\ii\tilde{G}_{k_{1}}^{\left(0\right)}\left(\omega\right)*\ldots*\ii\tilde{G}_{k_{2n+1}}^{\left(0\right)}\left(\omega\right),\label{eq:Sigma_2n+1_conv}
\end{align}
where $*$ denotes convolution. Eq.~(\ref{eq:Sigma_2n+1_conv}) shows
that the self-energy decomposes into a sum over baths of $2n+1$ photons,
where each many-body state of $2n+1$ photons contributes a Lorentzian
centered around $\sum_{i=1}^{2n+1}\omega_{k_{i}}$ with a width of
$\left(2n+1\right)\Ginb$. The weight of each Lorentzian in $\Sigma_{k;2n+1}^{\textrm{b}}\left(\omega\right)$
is determined by a product of factors $f_{k_{i}}^{2}$ of the modes
$k_{i}$ that form the $2n+1$-photons mode; since $f_{k}^{2}\sim1/\omega_{k}$,
the leading contribution originates from many-body states involving
many low-frequency modes.

\subsection{\label{subsec:cpb_fgr}Decay rate in the thermodynamic limit}

Consider the thermodynamic limit, $N\rightarrow\infty$, such that
$\Delta\rightarrow0$ and the modes of $\mathcal{H}_{0}$ form a continuum.
In that case, following the discussion in Section \ref{sec:toy-model},
the decay rate of a mode $k$ induced by the bath of $2n+1$ photons
is given by the imaginary part of the respective self-energy evaluated
at $\omega=\omega_{k}$, $\Gamma_{k;2n+1}^{\textrm{FGR}}=2\mathrm{Im}\Sigma_{k;2n+1}\left(\omega_{k}\right)$.
Using the unbroadened free propagators, $\ii\tilde{G}_{k}^{\left(0\right)}\left(t\right)=e^{-\ii\omega_{k}\left|t\right|}$,
we have
\begin{align}
\Gamma_{k;2n+1}^{\textrm{FGR}}= & \frac{2E_{J}^{2}f_{k}^{2}e^{-\sum_{k^\prime}f_{k^\prime}^{2}}}{\left(2n+1\right)!}\nonumber \\
 & \times\int_{-\infty}^{\infty}\id te^{\ii\omega_{k}t}\left[\sum_{k^{\prime}}f_{k^{\prime}}^{2}e^{-\ii\omega_{k^{\prime}}t}\right]^{2n+1}.\label{eq:GFGR_2n+1}
\end{align}
To illustrate the structure of $\Gamma_{k;2n+1}^{\textrm{FGR}}$,
it is useful to expand the brackets inside the integral and carry
out the integration over time. We arrive at
\begin{align}
\Gamma_{k;2n+1}^{\textrm{FGR}}= & \frac{2E_{J}^{2}f_{k}^{2}e^{-\sum_{k^\prime}f_{k^\prime}^{2}}}{\left(2n+1\right)!}\sum_{k_{1},\ldots,k_{2n+1}}\prod_{i=1}^{2n+1}f_{k_{i}}^{2}\nonumber \\
 & \times2\pi\delta\left(\omega_{k}-\omega_{k_{1}}-\ldots-\omega_{k_{2n+1}}\right),
\end{align}
which shows that the decay rate induced by $1\rightarrow2n+1$ decay
processes is given by a sum over all $2n+1$-photon states with energy
equal to $\omega_{k}$. In practice, it is more convenient to evaluate
$\Gamma_{k;2n+1}^{\textrm{FGR}}$ using Eq.~(\ref{eq:GFGR_2n+1});
however, Eq.~(\ref{eq:GFGR_2n+1}) involves a sum over the discrete
modes, which we would like to convert into an integral over the continuum
of modes in the thermodynamic limit, $\sum_{k}\rightarrow\int\id\omega^{\prime}/\Delta\left(\omega^{\prime}\right)$.
Since the integral is IR-divergent, $f_{k}^{2}\sim1/\omega_{k}$,
one must keep a finite IR cutoff, $\Delta/2$ (the lowest frequency
mode), that gives rise to a time-dependent factor, denoted here by
$\gamma_{1/2}\left(t\right)$, which is formally given by the Euler-Maclaurin
summation formula, and whose origin is similar to that of the Euler-Mascheroni
constant, $\gamma=\int_{1}^{\infty}\id x\left(1/\left\lfloor x\right\rfloor -1/x\right)\approx0.5772$
(where $\left\lfloor x\right\rfloor $ is the integer part of $x$):
\begin{equation}
\sum_{k^{\prime}}f_{k^{\prime}}^{2}e^{-\ii\omega_{k^{\prime}}t}\approx\int_{\Delta/2}^{\omega_{p}}\id\omega^{\prime}\frac{f^{2}\left(\omega^{\prime}\right)e^{-\ii\omega^{\prime}t}}{\Delta\left(\omega^{\prime}\right)}+2z\gamma_{1/2}\left(t\right).
\end{equation}
Note that $f_{k}^{2}\sim\Delta$, such that $f_{k}^{2}/\Delta$ is
finite in the limit $\Delta\rightarrow0$. The factor $\gamma_{1/2}\left(t\right)$
varies slowly on a scale of $1/\Delta$, so that in the thermodynamic
limit, $\Delta\rightarrow0$, it may be replaced by its value at $t=0$:
\begin{equation}
2z\gamma_{1/2}=\sum_{k^{\prime}}f_{k^{\prime}}^{2}-\int_{\Delta/2}^{\omega_{p}}\id\omega^{\prime}\frac{f^{2}\left(\omega^{\prime}\right)}{\Delta\left(\omega^{\prime}\right)}.
\end{equation}
The factor $\gamma_{1/2}$ depends weakly on $\Delta,\Gamma_{0},\omega_{p}$,
and, in the limit $\Delta\ll\Gamma_{0},\omega_{p}$, we may approximate
\begin{equation}
\gamma_{1/2}\approx\int_{0}^{\infty}\id x\left(\frac{1}{\left\lfloor x\right\rfloor +1/2}-\frac{1}{x+1/2}\right)\approx1.27.
\end{equation}
Plugging everything into $\Gamma_{k;2n+1}^{\textrm{FGR}}$, we find:
\begin{align}
\Gamma_{k;2n+1}^{\textrm{FGR}}= & \frac{2E_{J}^{2}f_{k}^{2}}{\left(2n+1\right)!}e^{-\int_{\Delta/2}^{\omega_{p}}\id\omega^{\prime}f^{2}\left(\omega^{\prime}\right)/\Delta\left(\omega^{\prime}\right)}e^{-2z\gamma_{1/2}}\nonumber \\
 & \hspace{-1.5cm}\times\int_{-\infty}^{\infty}\id te^{\ii\omega t}\left[\int_{\Delta/2}^{\omega_{p}}\frac{\id\omega^{\prime}}{\Delta\left(\omega^{\prime}\right)}f^{2}\left(\omega^{\prime}\right)e^{-\ii\omega^{\prime}t}+2z\gamma_{1/2}\right]^{2n+1}.\label{eq:GFGR_2n+1_th_lim}
\end{align}
Eq.~(\ref{eq:GFGR_2n+1_th_lim}) yields the FGR result for the decay
rate induced by the $2n+1$-photon bath, and may be evaluated numerically.
To gain analytical insight, it is useful to approximate $f^{2}\left(\omega^{\prime}\right)/\Delta\left(\omega^{\prime}\right)\approx2ze^{-\omega^{\prime}/\omega_{c}}/\omega^{\prime}$,
where $\omega_{c}$ is the cutoff frequency, given approximately by
$\omega_{c}\sim\min\left\{ \omega_{p},\Gamma_{0}\right\} $, and let
$\omega_{p}\rightarrow\infty$ for the upper integration limit of
$\int\id\omega^{\prime}$. This leads to
\begin{align}
\Gamma_{k;2n+1}^{\textrm{FGR}} & =\frac{2E_{J}^{2}f_{k}^{2}}{\left(2n+1\right)!}e^{-2z\int_{\Delta/2}^{\infty}\id\omega^{\prime}e^{-\omega^{\prime}/\omega_{c}}/\omega^{\prime}}e^{-2z\gamma_{1/2}}\nonumber \\
 & \hspace{-1.5cm}\times\int_{-\infty}^{\infty}\id te^{\ii\omega_{k}t}\left[2z\int_{\Delta/2}^{\infty}\frac{\id\omega^{\prime}}{\omega^{\prime}}e^{-\omega^{\prime}/\omega_{c}}e^{-\ii\omega^{\prime}t}+2z\gamma_{1/2}\right]^{2n+1},\label{eq:GFGR_2n+1_anal_apprx}
\end{align}
which is evaluated in Appendix \ref{app:FGR_2n+1_calc}. The leading
contribution reads $\Gamma_{k;2n+1}^{\textrm{FGR}}/\Delta\sim\Delta^{2z}\times1/\omega_{k}^{2}\times\log^{2n}\left(\omega_{k}/\Delta\right)$;
therefore, taking the thermodynamic limit, the decay rate of each
bath vanishes as power law in the system size, $N^{-2z}$, with some
logarithmic corrections. However, the total decay rate $\Gamma_{k}^{\textrm{FGR}}/\Delta=\sum_{n=1}^{\infty}\Gamma_{k;2n+1}^{\textrm{FGR}}/\Delta$
remains finite, and is given by a Luttinger-liquid power law, $\Gamma_{k}^{\textrm{FGR}}/\Delta\sim\omega_{k}^{2z-2}$.

\begin{figure}
\begin{centering}
\includegraphics[width=0.99\columnwidth]{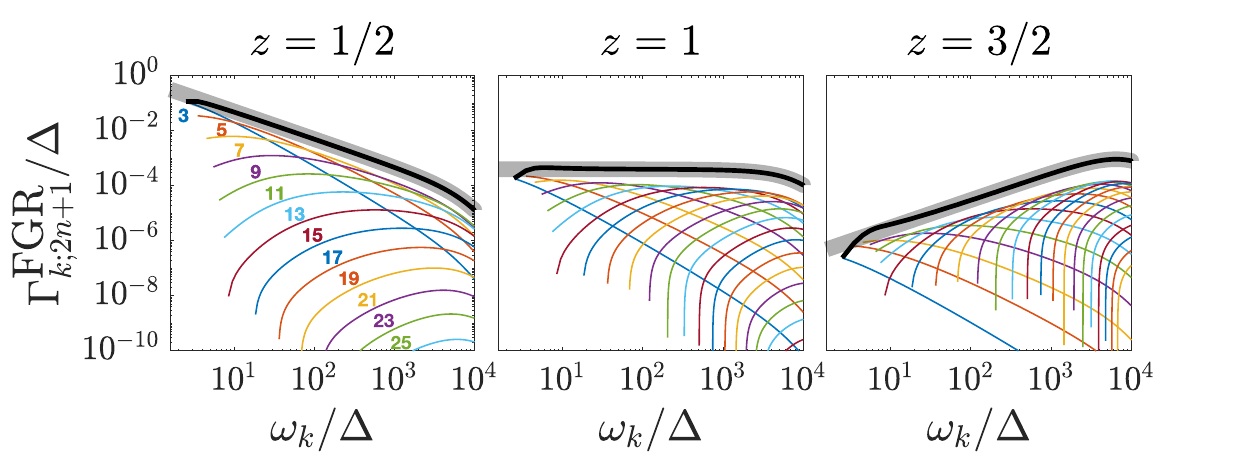}
\par\end{centering}
\caption{\label{fig:FGR_rates}Decay rates in the thermodynamic limit. The
colored lines depict the decay rates of the baths of $2n+1$ photons;
the decay rate of the $2n+1$ photons bath is plotted only for frequencies
satisfying $\omega_{k}>\left(2n+1\right)\Delta/2$. The bath indices
$N_{\textrm{out}}=2n+1$ are written explicitly next to the lines
in the $z=1/2$ panel, and the order of the lines is the same in the
$z=1$ and $z=3/2$ panels --- the blue line that starts on the top
left shows the 3-photon rate, the succeeding orange line shows the
5-photon rate, and so on. The black line is the sum of all rates,
$\Gamma_{k}^{\textrm{FGR}}=\sum_{n=1}^{\infty}\Gamma_{k;2n+1}^{\textrm{FGR}}$,
and the thick grey line is the expected result from Ref. \citep{kuzmin_observation_2023},
given by Eq.~(\ref{eq:GFGR_th}). The parameters used in this figure
are $\omega_{p}\rightarrow\infty$, $\Gamma_{0}/\Delta=10^{4}$, and
$E_{J}/\Delta=20,50,200$ for $z=1/2,1,3/2$, respectively.}
\end{figure}

The decay rates of the baths in the thermodynamic limit are shown
in Fig.~\ref{fig:FGR_rates}. The sum of rates agrees quantitatively
with the total decay rate,
\begin{equation}
\Gamma_{k}^{\textrm{FGR}}/\Delta=\frac{2\pi z}{\Gamma\left(2z\right)}\frac{E_{J}^{2}\omega_{k}^{2z-2}}{\omega_{c}^{2z}}e^{-2\omega_{k}/\omega_{c}},\label{eq:GFGR_th}
\end{equation}
which was derived in Ref. \citep{kuzmin_observation_2023}. Note that
for any $n$, the decay rate of the $2n+1$-photon bath $\Gamma_{k;2n+1}^{\textrm{FGR}}$
approaches 0 at large enough $\omega_{k}/\Delta$; this decrease is
countered by the emergence of new decay channels with higher $n$
as $\omega_{k}$ increases. It is interesting to note that at the
Schmid-Bulgadaev transition, $z=1$, the decrease in the rate of the
existing decay channels is exactly balanced by the emergence of the
new channels to produce a constant total decay rate at frequencies
$\Delta\ll\omega_{k}\ll\Gamma_{0}$.

\subsection{Decay rate at finite $N$}

\subsubsection{Approaching the thermodynamic limit}

Having obtained the FGR rates in the thermodynamic limit, we are now
interested in the decay rates for a finite array size $N$. First,
let us argue that, when $N$ is large enough (though finite) while
all other parameters are held fixed, Eq.~(\ref{eq:GFGR_2n+1_th_lim})
evaluated at finite $\Delta$ indeed corresponds to the decay rates
of the baths. This follows from the discussion of the toy model in
Section \ref{sec:toy-model}, where it was shown that, for external
broadening $\eta$ that is significantly larger than both the FGR
rate and the mode spacing (see Eq.~(\ref{eq:Pt_clean_summary})),
the decay rate is equal to the FGR rate. Considering the $2n+1$-photon
bath, its structure is revealed by Eq.~(\ref{eq:Sigma_2n+1_conv});
it is composed of many Lorentzians whose width is $\left(2n+1\right)\Ginb$,
in analogy to the external broadening $\eta$ of the modes in the
toy model. The mode spacing of $2n+1$-photon states is given by $\Delta_{2n+1}\left(\omega\right)=\left(2n\right)!\times\Delta^{2n+1}/\omega^{2n}$,
and the FGR rate scales as $\Gamma_{k;2n+1}^{\textrm{FGR}}\sim\Delta^{2z+1}/\omega_{k}^{2}$
up to logarithmic corrections, where $\Delta\sim1/N$. We thus find
that for fixed $\Ginb$ and any $\omega_{k}$, there is a crossover
scale for the array size, $N_{2n+1}^{\textrm{FGR}}$, defined by the
conditions of the latter case of Eq.~(\ref{eq:Pt_clean_summary}),
namely, $\Gamma_{k;2n+1}^{\textrm{FGR}}\ll\left(2n+1\right)\Ginb$
and $\Delta_{2n+1}\left(\omega_{k}\right)\ll\left(2n+1\right)\Ginb$.
Each of these inequalities defines a crossover scale, with $N_{2n+1}^{\textrm{FGR}}$
set by the larger of them; then, for $N\gg N_{2n+1}^{\textrm{FGR}}$,
we have $\Gamma_{k;2n+1}^{\textrm{in}}\rightarrow\Gamma_{k;2n+1}^{\textrm{FGR}}$.
$N_{2n+1}^{\textrm{FGR}}$ scales with frequency as $N_{2n+1}^{\textrm{FGR}}\sim\omega_{k}^{-\nu}$,
with $\nu=\max\left\{ 2n/\left(2n+1\right),2/\left(2z+1\right)\right\} $,
such that the FGR regime is reached faster (i.e. smaller $N$) at
larger frequencies, which can decay into a larger set of combinations of low-frequency photons.
Note that $N_{2n+1}^{\textrm{FGR}}$ depends on the order of the bath
$n$, such that different baths enter the FGR regimes at different
array sizes. This would be important in the following to understand
the cross-bath broadening.

Let us stress again that the order of limits is crucial; taking $\Ginb\rightarrow0$
and only then letting $N\rightarrow\infty$ would lead to a vanishing
decay rate. In the following, we show the convergence to the FGR rate
for a given bath is accelerated by the presence of the other baths,
using a self-consistent calculation.

\subsubsection{Self-consistent approach}

As discussed in Section \ref{sec:toy-model}, the crossover into the
FGR regime is determined by the ratio of the FGR decay rate and the
external broadening of the bath modes. It is evident from Eq.~(\ref{eq:Sigma_2n+1_conv})
that the bath of some mode $k$ is formed by many-body states involving
the other modes $k^{\prime}$, which are themselves broadened by nonlinear
scattering off the impurity. In order to incorporate these cascades
into the self-energy of the mode $k$, we replace the free propagators
$\tilde{G}_{k}^{\left(0\right)}$ in Eq.~(\ref{eq:Sigma_2n+1_bare})
with the dressed propagators $\tilde{G}_{k}^{\textrm{dr}}$:
\begin{align}
\Sigma_{k;2n+1}^{\textrm{dr}}\left(\omega\right)= & \frac{2\ii E_{J}^{2}f_{k}^{2}e^{-\sum_{k^{\prime}}f_{k^{\prime}}^{2}}}{\left(2n+1\right)!}\nonumber \\
 & \times\int_{-\infty}^{\infty}\id te^{\ii\omega t}\left[\sum_{k^{\prime}}\ii f_{k^{\prime}}^{2}\tilde{G}_{k^{\prime}}^{\textrm{dr}}\left(t\right)\right]^{2n+1}.\label{eq:Sigma_2n+1_dressed_def}
\end{align}
The dressed propagators $\tilde{G}_{k}^{\textrm{dr}}$ are calculated
with the dressed self-energies $\Sigma_{k;2n+1}^{\textrm{dr}}$,
\begin{equation}
\tilde{G}_{k}^{\textrm{dr}}\left(\omega\right)=\frac{2\omega_{k}}{\omega^{2}-\left[\omega_{k}-\left(\Sigma_{k}^{\textrm{dr}}\left(\omega\right)-\mathrm{Re}\Sigma_{k}^{\textrm{dr}}\left(\omega=0\right)\right)-\ii\Ginb/2\right]^{2}},
\end{equation}
where $\Sigma_{k}^{\textrm{dr}}=\sum_{n}\Sigma_{k;2n+1}^{\textrm{dr}}$.
Eq.~(\ref{eq:Sigma_2n+1_dressed_def}) thus defines a set of $N+1$
(one for each mode $k$) coupled integral equations, which are very
difficult to solve. This formidable task may be greatly simplified
by taking advantage of the structure of the self-energy in Eq.~(\ref{eq:Sigma_2n+1_conv}).
As mentioned above, each state of $2n+1$ photons contributes a Lorentzian
centered around $\sum_{i=1}^{2n+1}\omega_{k_{i}}$. In the evaluation
of the propagator $G_{k}$ of some mode $k$, we are only interested
in the self-energy $\Sigma_{k}$ at frequencies $\omega\sim\omega_{k}$;
the contribution of states with $\left|\sum_{i=1}^{2n+1}\omega_{k_{i}}-\omega_{k}\right|\gtrsim\Delta$
is negligible. In particular, states of $2n+1$ photons involving
modes $k^{\prime}\ge k$ can be discarded in the evaluation of $\Sigma_{k}$
--- namely,
\begin{align}
\Sigma_{k;2n+1}^{\textrm{dr}}\left(\omega\right)\approx & \frac{2\ii E_{J}^{2}f_{k}^{2}e^{-\sum_{k^{\prime}}f_{k^{\prime}}^{2}}}{\left(2n+1\right)!}\nonumber \\
 & \hspace{-1cm}\times\int_{-\infty}^{\infty}\id te^{\ii\omega t}\left[\sum_{k^{\prime}<k}\ii f_{k^{\prime}}^{2}\tilde{G}_{k^{\prime}}^{\textrm{dr}}\left(t\right)\right]^{2n+1}.\label{eq:Sigma_2n+1_dressed}
\end{align}
In other words, the self-energy $\Sigma_{k}^{\textrm{dr}}$ and propagator
$G_{k}^{\textrm{dr}}$ depend only on propagators at lower modes,
$k^{\prime}<k$, and the self-consistent equations may be solved iteratively
in a single run over the modes.

Let us note that the modes $k^{\prime}\ge k$ could affect the elastic
mode shift $\Delta\omega_{k}\sim\mathrm{Re}\Sigma_{k}\left(\omega_{k}\right)$
induced by the interaction Hamiltonian $\mathcal{H}_{I}$. We assume
that the contribution of the terms corresponding to the modes $k^{\prime}\ge k$
is negligible compared to the rest of the terms; we also introduce
some disorder in our numerical calculations (see below), which masks
the effect of the elastic phase shift.

The self-consistent approach accounts for the broadening of the propagators
induced by all of the baths. In order to highlight the interplay between
the different baths, we also apply a partial self-consistent approach,
in which we evaluate the self-energy
\begin{align}
\Sigma_{k;2n+1}^{\textrm{p}}\left(\omega\right)= & \frac{2\ii E_{J}^{2}f_{k}^{2}e^{-\sum_{k^{\prime}}f_{k^{\prime}}^{2}}}{\left(2n+1\right)!}\nonumber \\
 & \hspace{-1cm}\times\int_{-\infty}^{\infty}\id te^{\ii\omega t}\left[\sum_{k^{\prime}<k}\ii f_{k^{\prime}}^{2}\tilde{G}_{k^{\prime};2n+1}^{\textrm{p}}\left(t\right)\right]^{2n+1},\label{eq:Sigma_2n+1_partial}
\end{align}
where $\tilde{G}_{k^{\prime};2n+1}^{\textrm{p}}$ is the propagator
broadened only by the bath of $2n+1$ photons:
\begin{align}
\tilde{G}_{k^{\prime};2n+1}^{\textrm{p}}\left(\omega\right)= & 2\omega_{k}/\Bigg\{\omega^{2}-\Big[\omega_{k}-\ii\Ginb/2\nonumber \\
 & \hspace{-1.2cm}-\left(\Sigma_{k;2n+1}^{\textrm{p}}\left(\omega\right)-\mathrm{Re}\Sigma_{k;2n+1}^{\textrm{p}}\left(\omega=0\right)\right)\Big]^{2}\Bigg\}.
\end{align}
This partial approach incorporates the broadening that each bath imposes
on itself, but ignores inter-bath broadening.

\subsubsection{Numerical evaluation of the decay rates}

In the following, we evaluate the decay rate induced by a Cooper-pair
box impurity at finite array size $N$. We initialize all propagators
with some external broadening $\Ginb$, and evaluate the self-energy
using Eqs. (\ref{eq:Sigma_2n+1_bare}), (\ref{eq:Sigma_2n+1_dressed}),
or (\ref{eq:Sigma_2n+1_partial}). We plug the self-energies into
the propagator,
\begin{equation}
G_{k}\left(\omega\right)=\frac{4z\Delta\left(1-\left(\omega_{k}/\omega_{p}\right)^{2}\right)}{\omega^{2}-\left[\omega_{k}-\left(\Sigma_{k}\left(\omega\right)-\mathrm{Re}\Sigma_{k}\left(\omega=0\right)\right)-\ii\Ginb/2\right]^{2}}.\label{eq:Gkw}
\end{equation}
calculate the inverse Fourier transform to get $G_{k}\left(t\right)$,
and extract the decay rate $\Gamma_{k}^{\textrm{in}}$ from the late-time
dynamics by fitting $G_{k}\left(t\rightarrow\infty\right)$ to a decaying
exponential $e^{-\Gamma_{k}^{\textrm{in}}t}$.

We use experimentally-relevant parameters, specified in the caption
of Fig.~\ref{fig:cpb_rates}. Importantly, we use the normalized impedance
$z=1/2$, where the FGR decay rates decrease rapidly with the order
of the bath, as discussed in Section \ref{subsec:cpb_fgr} and illustrated
in Fig.~\ref{fig:FGR_rates}. We discard the contributions of baths
with $2n+1\ge11$, which are negligible. Note that it is crucial to
keep a finite plasma frequency $\omega_{p}$ in our calculations;
in the limit $\omega_{p}\rightarrow\infty$ and $v\rightarrow\infty$,
the dispersion relation is nearly linear, $\omega_{k}\approx vk$,
leading to a massive degeneracy of the many-body modes. In addition
to keeping $\omega_{p}$ finite, we introduce some disorder that spreads
out the many-body modes. Some details regarding the added disorder,
as well as a discussion of RG considerations at low-frequency modes,
are given in Appendix \ref{app:numerical_details}.

\subsubsection{Results and discussion}

\begin{figure*}
\begin{centering}
\includegraphics[height=0.3\textwidth]{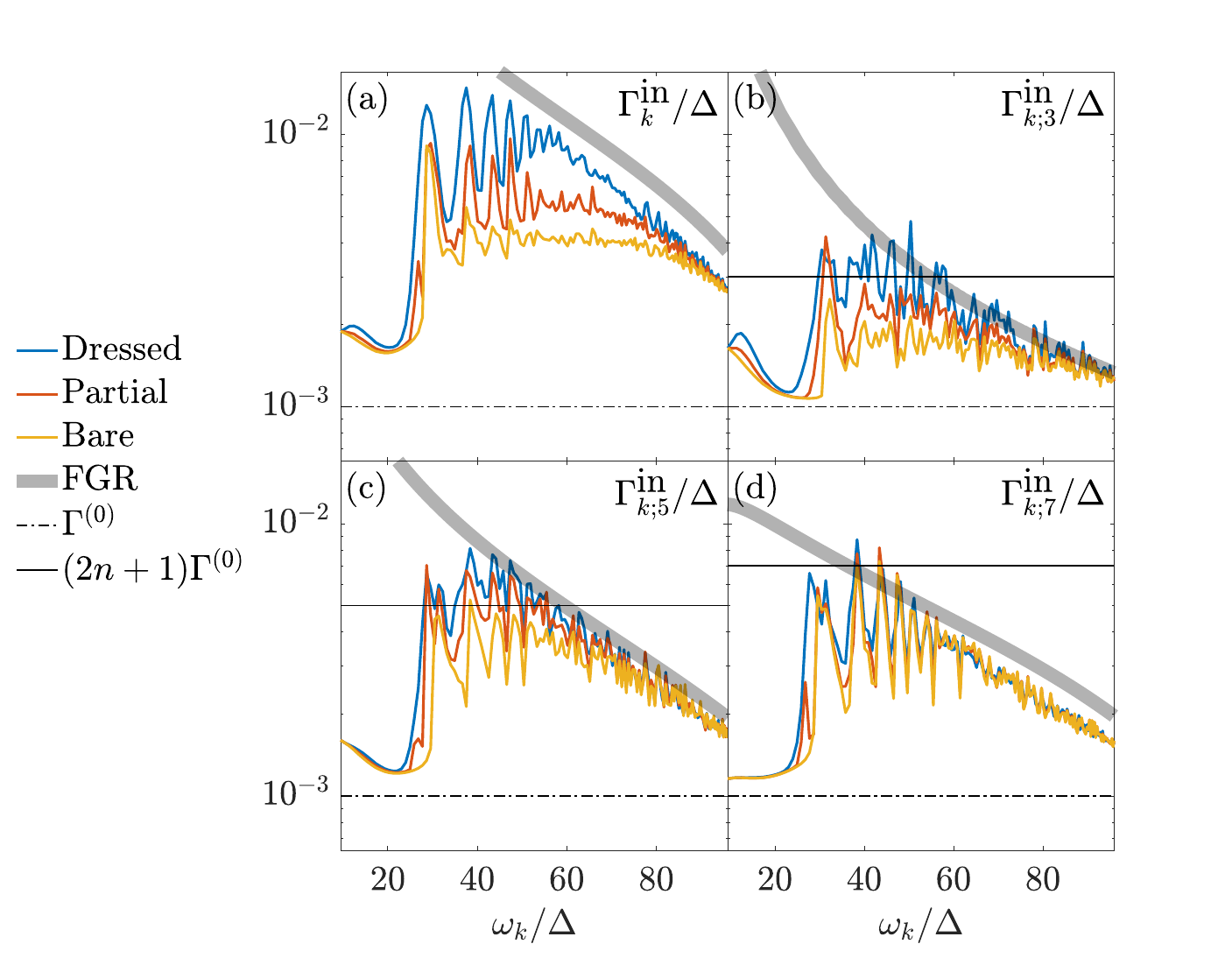}\hspace{-0.6cm}\includegraphics[height=0.3\textwidth]{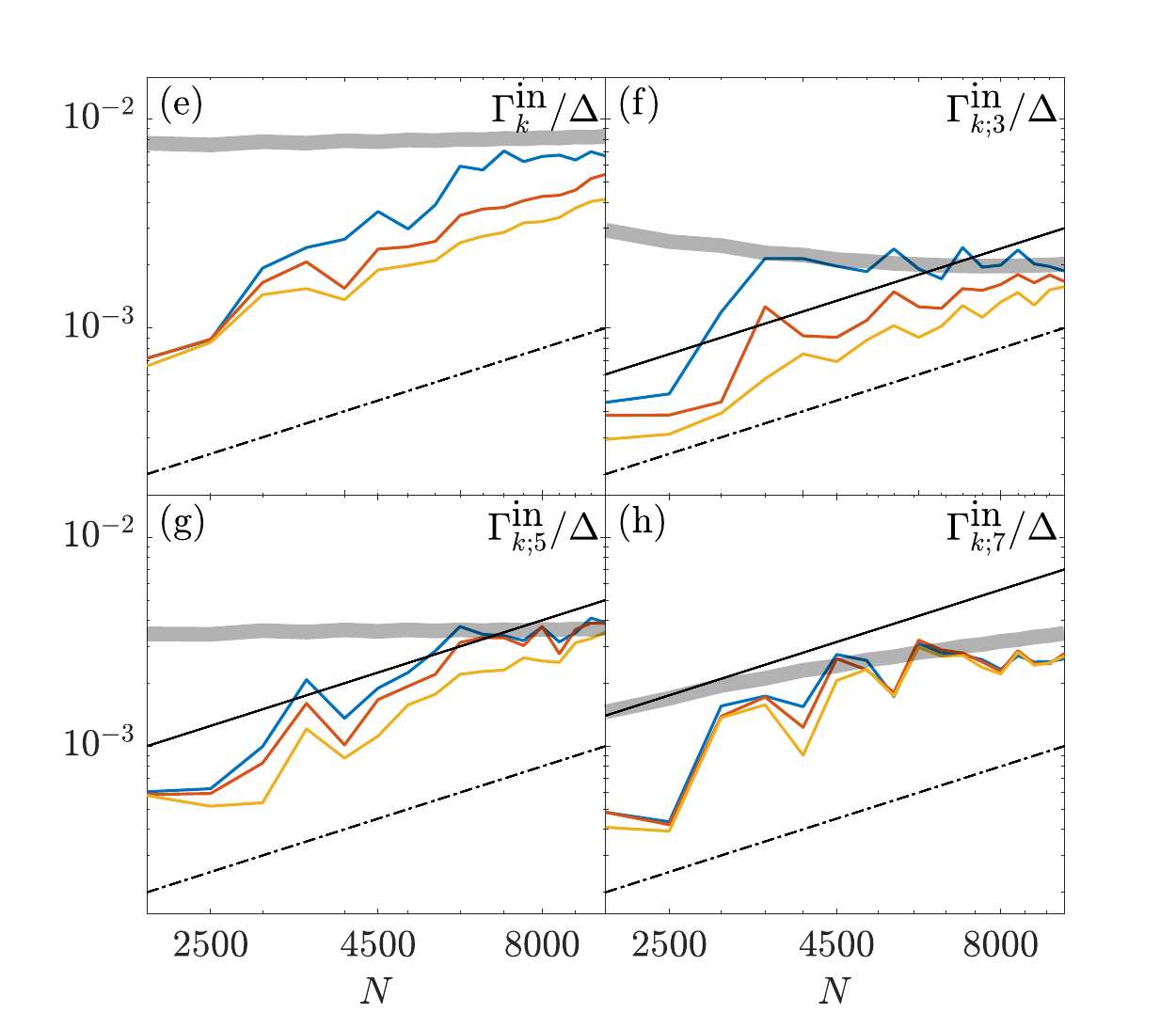}\hspace{-0.1cm}\includegraphics[height=0.3\textwidth]{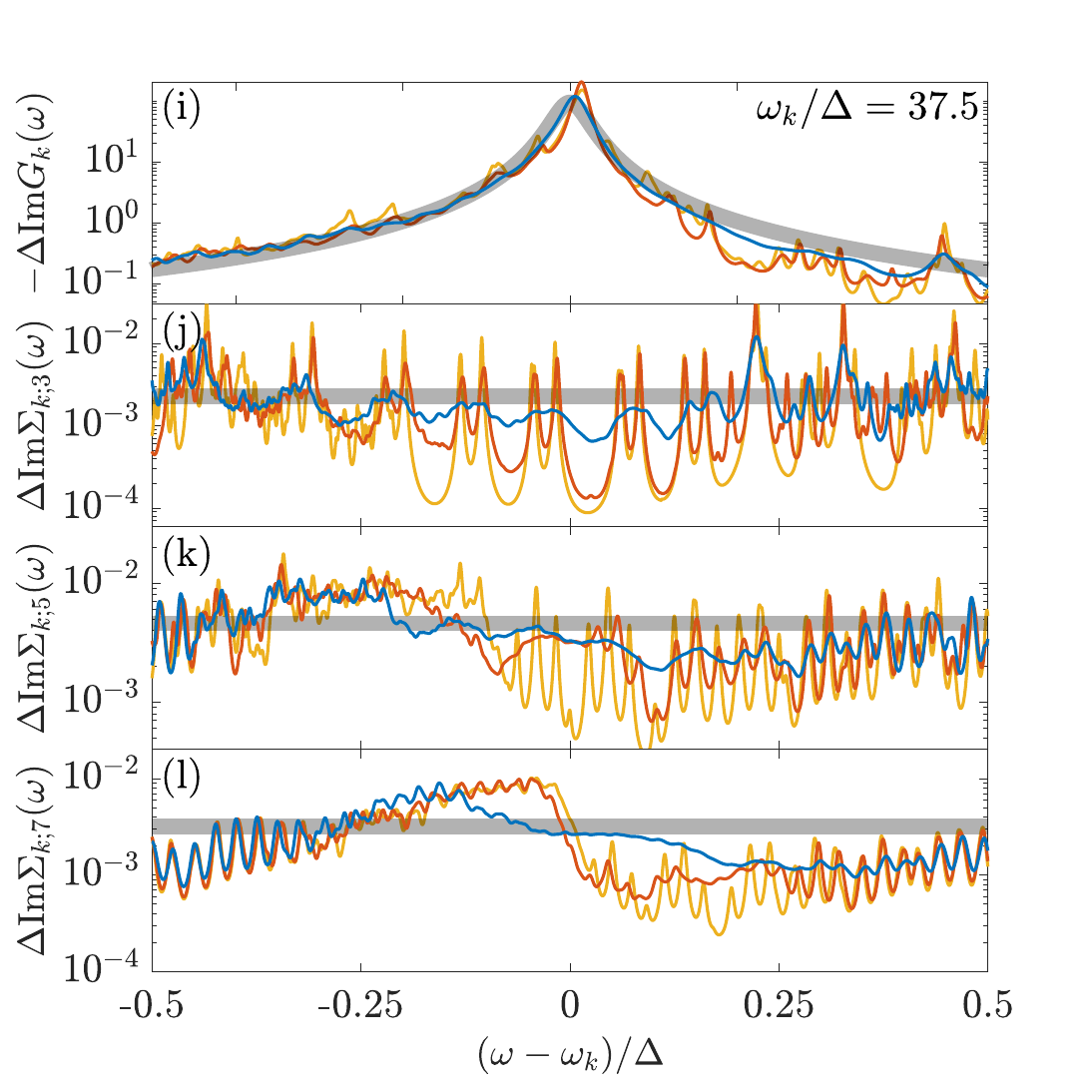}
\par\end{centering}
\caption{\label{fig:cpb_rates}(a)-(d) Decay rates as a function of frequency
extracted from the propagators in the time-domain, evaluated from
Eq.~(\ref{eq:Gkw}), with the self-energy given by Eqs. (\ref{eq:Sigma_2n+1_dressed})
(dressed), (\ref{eq:Sigma_2n+1_partial}) (partial), and (\ref{eq:Sigma_2n+1_bare})
(bare), for $N=10^{4}$. (a) Decay rate induced by the total self-energy
$\Sigma_{k}$. (b)-(d) Decay rates induced by the baths of 3-, 5-,
and 7-photon states. All decay rates were averaged over 50 realizations
of disorder. (e)-(h) Decay rates as a function of $N$ for $\omega_{k}=0.6\times\omega_{p}$.
(i)-(l) Spectral function and imaginary part of the self-energies
$\Sigma_{k;2n+1}$ of the mode at $\omega_{k}/\Delta=37.5$ with $N=10^{4}$,
for a single realization of disorder. In all plots, the thick grey
line shows the results predicted by the FGR. The experimentally-relevant
parameters used for this figure are $E_{J}=1$ GHz, $E_{C}=20$ GHz,
$z=1/2$, $\omega_{p}=20$ GHz, and $v=3.33$ THz (in units of the
intergrain spacing $a$), such that, at $N=10^{4}$, $\Delta=0.167$
GHz, and $\protect\Ginb/\Delta=10^{-3}$.}
\end{figure*}

The decay rates as a function of $\omega_{k}$ at fixed $N$, as well
as a function of $N$ at fixed $\omega_{k}$, are shown in Fig.~\ref{fig:cpb_rates}.
The decay rates induced by each of the baths, $\Gamma_{k;2n+1}^{\textrm{in}}$,
agree with the general result of Eq.~(\ref{eq:Pt_clean_summary})
derived for the toy model in the respective limits, and, in particular,
$\Gamma_{k;2n+1}^{\textrm{in}}\rightarrow\Gamma_{k;2n+1}^{\textrm{FGR}}$
at large enough $\omega_{k}/\Delta$ or $N$. At small $\omega_{k}/\Delta$
or $N$, the decay rate of the $2n+1$-photon bath is given by $\left(2n+1\right)\Ginb$.
Note that, for the 7-photon bath, we find that $\Gamma_{k;7}^{\textrm{in}}\sim\Gamma_{k;7}^{\textrm{FGR}}$
at most frequencies, since the bare broadening of the 7-photon modes
is larger than the FGR rate, $\Gamma_{k;7}^{\textrm{FGR}}<7\Ginb$,
at most frequencies. A similar result is obtained for the 9-photon
bath, not shown in Fig.~\ref{fig:cpb_rates}. The oscillations of
the decay rates with frequency at small $\omega_{k}/\Delta$ are due
to the convexity of the dispersion relation, $\omega_{k}\approx vk/\sqrt{1+\left(vk/\omega_{p}\right)^{2}}$,
which shifts the energies of the many-body states with respect to
those of the single-photon states, as illustrated in Fig.~\ref{fig:intro_fig}.
The 3- and 5-photon baths, on the other hand, display a crossover
between the two regimes.

The effect of the cascade decay processes is clearly illustrated by
the crossover region between small and large $\omega_{k}/\Delta$
or $N$. As expected, the dressed propagator yields a larger decay
rate than that of the bare propagator. The dressing accelerates the
convergence of the decay rates to their FGR values with either frequency
or array size, and, importantly, enhances the maximally-allowed decay
rate induced by each bath. In Section \ref{sec:toy-model}, we showed
that the decay rate induced by a bath is necessarily bounded by the
largest broadening of the bath modes, such that, when the bare propagator
$G_{k}^{\textrm{b}}$ is considered (that is, excluding cascade processes),
it must hold that $\Gamma_{k;2n+1}^{\textrm{in}}<\left(2n+1\right)\Ginb$.
However, the dressing broadens the bath states; therefore, the decay
rate of the $2n+1$-photon bath extracted from the dressed propagator
may in fact exceed $\left(2n+1\right)\Ginb$.

While the accelerated crossover to the FGR regime and the enhancement
of the maximally-allowed decay rates occur for both partial ($G_{k}^{\textrm{p}}$)
and full ($G_{k}^{\textrm{dr}}$) dressing, these effects are particularly
prominent in the latter case, as demonstrated by Fig.~\ref{fig:cpb_rates}.
That is, the decay rate into the $n$-photon bath is enhanced by the
presence of $m\neq n$-photon baths. This is due to the different
convergence rates of the baths to their FGR regimes; as discussed
above, each bath is characterized by a crossover scale for the array
size $N_{2n+1}^{\textrm{FGR}}$, which depends on the bath order $2n+1$.
For a Cooper-pair box impurity with $z=1/2$, we find that baths formed
by many-photon states enter the FGR regime and induce a large decay
rate faster than baths of few-photon states. This is most evident
for the 3-photon bath --- there is a range of modes that are not
in the FGR regime of the 3-photon bath (i.e. $\Gamma_{k;3}^{\textrm{in}}<\Gamma_{k;3}^{\textrm{FGR}}$),
which are broadened by the 5-, 7-, and 9-photon baths, such that the
single-photon modes in this range are broadened beyond $\Ginb$. Therefore,
when a mode $k$ in this range decays into 3-photon states involving
another mode in this range ($\omega_{k}\rightarrow\omega_{k}+\omega_{k^{\prime\prime}}+\omega_{k^{\prime\prime\prime}}$,
where $k^{\prime}\lesssim k$ and $k^{\prime\prime},k^{\prime\prime\prime}$
are low frequency modes), the total broadening of the outgoing 3-photon
state exceeds $3\Ginb$, and therefore $\Gamma_{k;3}^{\textrm{in}}>3\Ginb$
is allowed. This effect accelerates the convergence of $\Gamma_{k;3}^{\textrm{in}}$
to its FGR value; the condition for the FGR to apply is then $\Gamma_{k;3}^{\textrm{FGR}}<\bar{\eta}_{3}$,
where $\bar{\eta}_{3}$ is the average width of the 3-photon states,
which takes into account the broadening induced by the cascade decay
processes.

\begin{figure}
\begin{centering}
\includegraphics[width=0.98\columnwidth]{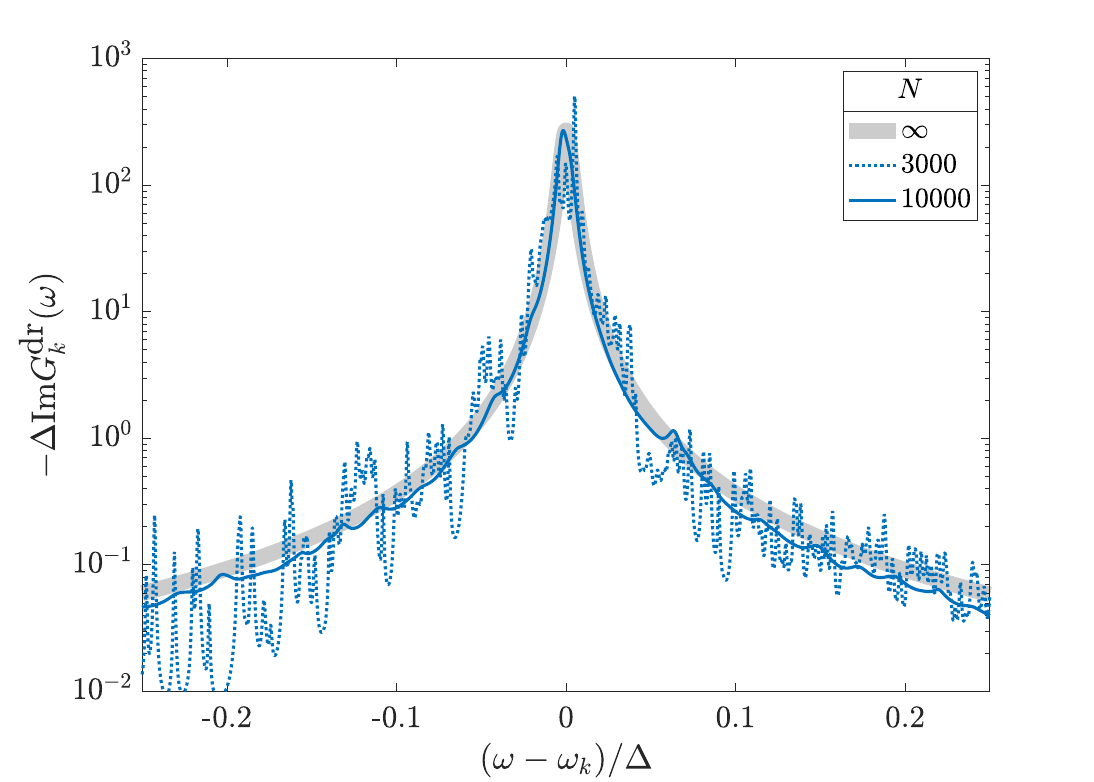}
\par\end{centering}
\caption{\label{fig:cpb_Gk}Spectral function extracted from the fully-dressed
propagator of the mode at $\omega_{k}=0.6\times\omega_{p}$ for two
values of the array size $N$, each for a single realization of disorder.
The thick grey line shows the spectral function for a semi-infinite
array. The parameters used are the same as in Fig.~\ref{fig:cpb_rates}.}
\end{figure}

Fig.~\ref{fig:cpb_rates} also shows the self-energies and the propagator
of one of the modes at some given array size $N$. The sharp features
of the propagator $G_{k}\left(\omega\right)$ are due to level repulsion
between the single-photon mode at $\omega_{k}$ and the many-body
states at comparable frequencies, which appear in the self-energies.
The fully-dressed propagator $G_{k}^{\textrm{dr}}$ and self-energies
$\Sigma_{k;2n+1}^{\textrm{dr}}$ are generally smoother than their
bare and partially-dressed counterparts (as most clearly illustrated
by the difference between $\Sigma_{k;3}^{\textrm{dr}}$ and $\Sigma_{k;3}^{\textrm{b}},\Sigma_{k;3}^{\textrm{p}}$),
since the Lorentzians in the dressed self-energies (see Eq.~(\ref{eq:Sigma_2n+1_conv}))
are smeared out by the broadening of the single-photon states comprising
the multi-photon states. Fig.~\ref{fig:cpb_Gk} further shows the
spectral function of the fully-dressed propagator for two different
array sizes at some given frequency --- as expected, level repulsion
is a lot more prominent for shorter array size.

The level repulsion in $G_{k}\left(\omega\right)$ may be observed
in a spectroscopy experiment, and was reported in Ref. \citep{mehta_down-conversion_2023}
for a fluxonium impurity, where the self-energies of the single-photon
modes are dominated by the contributions of the 2- and 3-photon baths.
Our calculation provides a prediction for the spectral function probed
in such an experiment, and could be applied to any nonlinearity (located
either at the boundary or in the bulk) with any number of multi-photon
baths. Importantly, let us note that the absence of anticrossings
in the spectroscopy measurement does not necessarily imply that the
single-photon decoheres with the FGR rate. Its coherence time could
be longer than its apparent lifetime extracted from the width of the
Lorentzian, if the condition $\bar{\eta}\gg\Gamma_{k}^{\textrm{FGR}}$
is not met, where $\bar{\eta}$ is the average width of the multi-photon
levels that couple to the single-photon state. Indeed, if $\Gamma_{k}^{\textrm{FGR}}\gtrsim\bar{\eta}$
and $\Delta_{n}\ll\bar{\eta}$ for the multi-photon mode spacings
$\Delta_{n}$, the propagator $G_{k}\left(\omega\right)$ would form
a smooth Lorentzian with a width $\Gamma_{k}^{\textrm{FGR}}$, yet
revival processes at large times would still occur, and the long-time
decoherence rate $\Gamma$ would be smaller than $\Gamma_{k}^{\textrm{FGR}}$.

\section{\label{sec:conclusions}Conclusions}

In this work, we showed how the FGR emerges in the decay of a single-photon
when the system approaches the thermodynamic limit. Keeping a finite
bare decoherence rate, necessary to obtain an exponential decay of
the single-photon states in the long-time limit for any finite system
size, we derived the conditions for the rates induced by the multi-photon
baths to reach the values predicted by the FGR. We calculated the
self-energies of the single-photon propagators in a self-consistent
manner, and showed that the crossover of the $n$-photon bath into
the FGR regime is accelerated by the presence of the other $m\neq n$-photon
baths, due to cascade decay processes into low-frequency modes.

This work provides a framework to analyze the experiment in Ref. \citep{mehta_down-conversion_2023},
and allows one to reproduce the measured single-photon spectroscopy
picture. The level repulsion between single- and multi-photon states
reported in Ref. \citep{mehta_down-conversion_2023} is clearly visible
in Fig.~\ref{fig:cpb_rates}, and our calculations should apply to
any bath with any number of photons, and any type of weak bulk or
boundary nonlinearity. As discussed above, the absence of level repulsions
does not necessarily imply that the single-photon decoheres with the
FGR rate, since the decoherence rate could still be limited by the
bare decoherence rate of the array modes, which is unrelated to the
bulk or boundary nonlinearity. In the context of quantum information
processing, this effect could be leveraged to engineer a system of
a finite multimode resonator superstrongly coupled to a qubit to harvest
such revival processes and prolong the lifetime of the single-photon
and qubit states.

As mentioned above, the emergence of the FGR is closely related to
questions about thermalization in large many-body systems, Fock-space
delocalization, and the onset of chaos \citep{altshuler_quasiparticle_1997,bulchandani_onset_2022, crowley_mean_2022},
which have also been discussed recently in the context of circuit
quantum electrodynamics \citep{pekola_thermalization_2024}. Let us
note that in this work we considered a Cooper-pair box impurity, which,
upon taking the scaling limit $E_{J},\omega_{c}\rightarrow\infty$
(where $\omega_{c}\sim\min\left\{ \Gamma_{0},\omega_{p}\right\} $)
with $E_{J}^{\star}=\left(E_{J}/\omega_{c}^{z}\right)^{1/\left(1-z\right)}$
fixed (for $z<1$), leads to the boundary sine-Gordon model, which
is integrable \citep{ghoshal_boundary_1994}. Our results imply that
the FGR should still emerge in the boundary sine-Gordon model, despite
of its integrability, with the exact inelastic decay rates in the
thermodynamic limit calculated in Ref. \citep{burshtein_inelastic_2024}.
The emergence of inelastic decay in integrable systems raises an analogous
question in the context of thermalization. Ref. \citep{pekola_thermalization_2024}
demonstrated that a closed system with a generic three-wave mixing
term, initialized in some excited state, relaxes to a Bose-Einstein
distribution of photons. While it is well-known that integrable systems
thermalize into generalized Gibbs ensembles \citep{rigol_relaxation_2007},
could the cosine nonlinearity of the (closed) boundary sine-Gordon
model still give rise to a standard Bose-Einstein distribution of
the photons (which are not the eigenstates of the Hamiltonian)? And
if so, how is the convergence rate to this distribution related to
the FGR? We leave these questions as a future perspective.

\subsubsection*{Acknowledgements}

We would like to thank Ehud Altman, Boris Altshuler, Anushya Chandran, Roman Kuzmin,
and Maxim Vavilov for fruitful discussions. Our work has been supported
by the Israel Science Foundation (ISF) and the Directorate for Defense
Research and Development (DDR\&D) through Grant No. 3427/21, the ISF
grant No. 1113/23, and the US-Israel Binational Science Foundation
(BSF) through Grant No. 2020072. A.B. is also supported by the Adams
Fellowship Program of the Israel Academy of Sciences and Humanities.

\nocite{burshtein_2025_15319197}

\appendix

\section{\label{app:FGR_2n+1_calc}Calculation of $\Gamma_{k;2n+1}^{\textrm{FGR}}$}

Consider Eq.~(\ref{eq:GFGR_2n+1_anal_apprx}). The integral over $\omega^{\prime}$
evaluates as
\begin{equation}
\int_{\Delta/2}^{\infty}\frac{\id\omega^{\prime}}{\omega^{\prime}}e^{-\omega^{\prime}/\omega_{c}}e^{-\ii\omega^{\prime}t}=\Gamma\left(0,\frac{\Delta}{2\omega_{c}}+\frac{\ii\Delta t}{2}\right),
\end{equation}
where $\Gamma\left(s,x\right)$ is the incomplete gamma function \citep{abramowitz_handbook_1965}.
Plugging into Eq.~(\ref{eq:GFGR_2n+1_anal_apprx}), we find

\begin{widetext}

\begin{equation}
\Gamma_{k;2n+1}^{\textrm{FGR}}=2E_{J}^{2}f_{k}^{2}e^{-\sum_{k^\prime}f_{k^\prime}^{2}}\sum_{m=0}^{2n+1}\frac{\left(2z\right)^{2n+1}\gamma_{1/2}^{2n+1-m}}{\left(2n+1\right)!}{2n+1 \choose m}\int_{-\infty}^{\infty}\id te^{\ii\omega_{k}t}\Gamma^{m}\left(0,\frac{\Delta}{2\omega_{c}}+\frac{\ii\Delta t}{2}\right).\label{eq:GFGR_2n+1_calc_1}
\end{equation}
The integral over $t$ may be evaluated by choosing a contour that
goes around the branch cut of the incomplete gamma function, yielding
\begin{equation}
\int_{-\infty}^{\infty}\id te^{\ii\omega_{k}t}\Gamma^{m}\left(0,\frac{\Delta}{2\omega_{c}}+\frac{\ii\Delta t}{2}\right)=\frac{\ii e^{-\omega_{k}/\omega_{c}}}{\Delta/2}\int_{0}^{\infty}\id xe^{-x\omega_{k}/\left(\Delta/2\right)}\left[\left(\Gamma\left(0,-x\right)\right)^{m}-\left(\Gamma\left(0,-x\right)+2\pi\ii\right)^{m}\right],
\end{equation}
where $\Gamma\left(0,-x\right)\equiv\Gamma\left(0,xe^{i\pi-i\delta}\right)$.
In order to evaluate the remaining integral analytically, we replace
the incomplete gamma function with its series expansions at small
and large $x$, $\Gamma\left(0,-x\right)=f\left(x\right)-\ii\pi$,
with $f\left(x\ll1\right)\approx-\log x-\gamma$ (where $\gamma\approx0.5772$
is the Euler-Mascheroni constant), and $f\left(x\gg1\right)\sim-e^{x}/x$,
and find
\begin{equation}
\int_{-\infty}^{\infty}\id te^{\ii\omega_{k}t}\Gamma^{m}\left(0,\frac{\Delta}{2\omega_{c}}+\frac{\ii\Delta t}{2}\right)\approx\frac{2e^{-\omega_{k}/\omega_{c}}}{\Delta/2}\sum_{l=0}^{\left\lfloor \left(m-1\right)/2\right\rfloor }{m \choose 2l+1}\left(-1\right)^{l}\pi^{2l+1}\int_{0}^{\infty}\id xe^{-x\omega_{k}/\left(\Delta/2\right)}f^{m-2l-1}\left(x\right).
\end{equation}
Note that the integrand decays at large $x$ provided that $\omega_{k}>n\Delta$,
which is a necessary condition to allow for a $1\rightarrow2n+1$
decay process. Indeed, for large enough $\omega_{k}/\Delta$, the
integrand decays very rapidly with $x$, and one may use the expansion
of $f\left(x\ll1\right)$, leading to
\begin{align}
\left(-1\right)^{m-2l-1}\int_{0}^{\infty}\id xe^{-x\omega_{k}/\left(\Delta/2\right)}\left(\log x+\gamma\right)^{m-2l-1}\nonumber \\
 & \hspace{-4cm}=\frac{\Delta/2}{\omega_{k}}\sum_{r=0}^{m-2l-1}\sum_{p=0}^{r}{m-2l-1 \choose r}{r \choose p}\left(-1\right)^{m+p-1}\gamma^{m-2l-r-1}\Gamma^{\left(r-p\right)}\left(1\right)\log^{p}\left(\frac{\omega_{k}}{\Delta/2}\right),
\end{align}
where $\Gamma^{\left(s\right)}\left(x\right)$ is the $s$ derivative
of the gamma function. Finally, the exponential factor $e^{-\sum_{k^\prime}f_{k^\prime}^{2}}$
may also be evaluated analytically using $f_{k}^{2}\approx2z\Delta e^{-\omega_{k}/\omega_{c}}/\omega_{k}$:
\begin{equation}
e^{-\sum_{k^\prime}f_{k^\prime}^{2}}\approx\left(\frac{\Delta/2}{\omega_{c}}\right)^{2z}e^{2z\left(\gamma-\gamma_{1/2}\right)}.
\end{equation}
Plugging everything back into Eq.~(\ref{eq:GFGR_2n+1_calc_1}), we find
\begin{align}
\Gamma_{k;2n+1}^{\textrm{FGR}}/\Delta= & \frac{2\left(2z\right)^{2n+2}E_{J}^{2}e^{-2\omega_{k}/\omega_{c}}}{\omega_{k}^{2}}\left(\frac{\Delta/2}{\omega_{c}}\right)^{2z}e^{2z\left(\gamma-\gamma_{1/2}\right)}\nonumber \\
 & \times\sum_{m=0}^{2n+1}\sum_{l=0}^{\left\lfloor \left(m-1\right)/2\right\rfloor }\sum_{r=0}^{m-2l-1}\sum_{p=0}^{r}{2n+1 \choose m}{m \choose 2l+1}{m-2l-1 \choose r}{r \choose p}\frac{1}{\left(2n+1\right)!}\nonumber \\
 & \times\left(-1\right)^{m+l+p-1}\gamma_{1/2}^{2n+1-m}\gamma^{m-2l-r-1}\pi^{2l+1}\Gamma^{\left(r-p\right)}\left(1\right)\log^{p}\left(\frac{\omega_{k}}{\Delta/2}\right).\label{eq:GFGR_2n+1_calc_2}
\end{align}
Keeping only the leading logarithmic correction at $\Delta\ll\omega_{k}\ll\omega_{c}$
from the sums above is enough to show that the total decay rate, $\Gamma_{k}^{\textrm{FGR}}=\sum_{n=1}^{\infty}\Gamma_{k;2n+1}^{\textrm{FGR}}$,
recovers the expected Luttinger-liquid power law: The leading term
in Eq.~(\ref{eq:GFGR_2n+1_calc_2}) corresponds to $m=2n+1$, $l=0$,
$r=2n$, and $p=2n$, and reads (up to the decaying exponential $e^{-2\omega_{k}/\omega_{c}}$
and numerical factors that depend on $z$)
\begin{equation}
\Gamma_{k;2n+1}^{\textrm{FGR}}/\Delta\sim\frac{E_{J}^{2}}{\omega_{k}^{2}}\left(\frac{\Delta}{\omega_{c}}\right)^{2z}\times\frac{\left(2z\right)^{2n}}{\left(2n\right)!}\log^{2n}\left(\frac{\omega_{k}}{\Delta/2}\right),
\end{equation}
leading to
\begin{equation}
\Gamma_{k}^{\textrm{FGR}}/\Delta\sim\frac{E_{J}^{2}}{\omega_{k}^{2}}\left(\frac{\Delta}{\omega_{c}}\right)^{2z}\left[\cosh\left(\log\left(\left(\frac{\omega_{k}}{\Delta/2}\right)^{2z}\right)\right)-1\right]\sim E_{J}^{2}\omega_{c}^{-2z}\omega_{k}^{2z-2}.
\end{equation}
To recover the full numerical prefactor of $\Gamma_{k}^{\textrm{FGR}}$
it is necessary to collect all logarithmic corrections of the same
order in $\Gamma_{k}^{\textrm{FGR}}=\sum_{n=1}^{\infty}\Gamma_{k;2n+1}^{\textrm{FGR}}$
from the cumbersome expression in Eq.~(\ref{eq:GFGR_2n+1_calc_2}).
This is verified numerically in Fig.~\ref{fig:FGR_rates}, and easily
follows from Eq.~(\ref{eq:GFGR_2n+1}), since
\begin{align}
\sum_{n=1}^{\infty}\Gamma_{k;2n+1}^{\textrm{FGR}}= & 2E_{J}^{2}f_{k}^{2}e^{-\sum_{k^\prime}f_{k^\prime}^{2}}\int_{-\infty}^{\infty}\id te^{\ii\omega_{k}t}\sum_{n=1}^{\infty}\frac{1}{\left(2n+1\right)!}\left[\sum_{k^{\prime}}f_{k^{\prime}}^{2}e^{-\ii\omega_{k^{\prime}}t}\right]^{2n+1}\nonumber \\
= & 2E_{J}^{2}f_{k}^{2}e^{-\sum_{k^\prime}f_{k^\prime}^{2}}\int_{-\infty}^{\infty}\id te^{\ii\omega_{k}t}\left[\sinh\left(\sum_{k^{\prime}}f_{k^{\prime}}^{2}e^{-\ii\omega_{k^{\prime}}t}\right)-1\right]\nonumber \\
\approx & E_{J}^{2}f_{k}^{2}\int_{-\infty}^{\infty}\id te^{\ii\omega_{k}t}\exp\left(-\sum_{k^{\prime}}f_{k^{\prime}}^{2}\left(1-e^{-\ii\omega_{k^{\prime}}t}\right)\right),
\end{align}
where, going from the second to the third line above, we used $e^{-\sum_{k^\prime}f_{k^\prime}^{2}}\sim\Delta^{2z}$
to discard terms that vanish in the thermodynamic limit. We thus recover
the total decay rate in the thermodynamic limit, which was evaluated
in Ref. \citep{kuzmin_observation_2023}.

\end{widetext}

\section{\label{app:numerical_details}Details of numerical evaluation of
the decay rates}

\subsection{Average over disorder}

As mentioned above, one must keep a finite plasma frequency in order
to break the massive degeneracy of the many-body modes. This degeneracy
is further avoided by introducing some disorder that is always present
in any experimental setup, due to deviations in the values of the
array Josephson couplings and capacitances. We assume a realistic
deviation of $\pm10\%$ around the nominal values of the line parameters
$E_{J}^{\textrm{line}},C^{\textrm{line}},C_{g}$ in Eq.~(\ref{eq:H_CPB}),
and, for each realization, find the modes by solving the corresponding
generalized eigenvalue problem numerically. An example for the modes
of 1-, 3-, and 5-photon states for a single realization of disorder
is shown in Fig.~\ref{fig:intro_fig}. In the calculation of the dressed
propagators, we use each realization of modes to calculate the propagators
of the modes iteratively, going up from the low-frequency modes. The
spectral functions measured in an experiment correspond to single
realizations of disorder, as the ones displayed in Figs. \ref{fig:cpb_rates}
and \ref{fig:cpb_Gk}. The decay rates shown in Fig.~\ref{fig:cpb_rates}
are averaged over 50 realizations of disorder.

\subsection{Renormalization group considerations at low frequencies}

In Figs. \ref{fig:cpb_rates} and \ref{fig:cpb_Gk}, we use $z=1/2$.
Note that since $z<1$, the boundary cosine operator $\mathcal{H}_{I}=E_{J}\cos\phi_{0}$
is relevant in an RG sense and cannot be treated as a perturbation
below the RG scale $E_{J}^{\star}=\left(E_{J}/\omega_{c}^{z}\right)^{1/\left(1-z\right)}$.
Using our choice of numerical parameters, we find $E_{J}^{\star}/\Delta<1$
at $N=10^{4}$ (and smaller at smaller array sizes), such that our
perturbative analysis is valid from the very first mode. Note that
in principle, if the RG scale were larger with several array modes
below it, it would have been possible to extend our treatment to these
low-frequency modes as well by expanding around the strong-coupling
fixed-point. A strong-coupling expansion should lead to the same structure
of the self-energy, but with new $f_{k}$ factors. It is then possible
to stitch together the two regimes, by using the $f_{k}$ factors
from the strong-coupling expansion at frequencies below $E_{J}^{\star}$,
and the perturbative $f_{k}$ at frequencies above $E_{J}^{\star}$.

\bibliography{FGR_bib}

\end{document}